\documentclass[lettersize,journal]{IEEEtran}
\usepackage{graphicx}
\usepackage{subcaption}
\usepackage{amsmath,amsfonts}
\usepackage{algorithmic}
\usepackage{algorithm}
\usepackage{array}
\usepackage{textcomp}
\usepackage{stfloats}
\usepackage{url}
\usepackage{verbatim}
\usepackage{graphicx}
\usepackage{cite}
\usepackage{hyperref}
\usepackage{cuted}
\usepackage{adjustbox}
\hypersetup{
  colorlinks=true,
  linkcolor=black,
  filecolor=black,      
  urlcolor=black,
  citecolor=black,
}

\usepackage{booktabs}
\usepackage{multirow}

\begin{document}

\newcommand\revision[1]{\textcolor{black}{#1}}
\newcommand\revisionsecond[1]{\textcolor{black}{#1}}

\title{Understanding and Predicting Temporal Visual Attention Influenced by Dynamic Highlights in Monitoring Task}
\author{Zekun Wu and Anna Maria Feit
\thanks{Zekun Wu and Anna Maria Feit are with Saarland University, Saarland Informatics Campus, Germany (e-mail: wuzekun$@$cs.uni-saarland.de; feit$@$cs.uni-saarland.de).}%
\thanks{This project is funded by Deutsche Forschungsgemeinschaft (DFG, German Research Foundation) -- project number 389792660 -- TRR 248 -- CPEC, see \url{https://perspicuous-computing.science}}%
}



\maketitle

\begin{abstract}
Monitoring interfaces are crucial for dynamic, high-stakes tasks where effective user attention is essential. Visual highlights can guide attention effectively but may also introduce unintended disruptions. To investigate this,  we examined how visual highlights affect users' gaze behavior in a drone monitoring task, focusing on when, how long, and how much attention they draw. We found that highlighted areas exhibit distinct temporal characteristics compared to non-highlighted ones, quantified using normalized saliency (NS) metrics. We found that highlights elicited immediate responses, with NS peaking quickly, but this shift came at the cost of reduced search efforts elsewhere, potentially impacting situational awareness. To predict these dynamic changes and support interface design, we developed the Highlight-Informed Saliency Model (HISM), which provides granular predictions of NS over time. These predictions enable evaluations of highlight effectiveness and inform the optimal timing and deployment of highlights in future monitoring interface designs, particularly for time-sensitive tasks.  
\end{abstract}


\section{Introduction}

\revision{In high-stakes complex monitoring environments like air traffic control and network operations centers, visual highlights are widely used to direct attention, highlight critical information, and enhance task performance~\cite{Kern2010,yang2017method,de2009towards,li2020augmented,lalle2019gaze,gingerich2015constructing}. For instance, dynamic color changes in monitoring dashboards—such as those used in flight displays or drone operations—can effectively draw attention to critical trends, helping users promptly address safety-critical situations\cite{li2020augmented, wu2024enhancing}. While intuitively effective, the precise influence of such visual highlights on user attention—how quickly they are noticed, how long they hold attention, or how they impact awareness of other interface elements—remains poorly understood and difficult to predict during design.}

\revision{Currently, deep neural networks are commonly used to predict user visual attention on GUIs~\cite{jiang2023ueyes, leiva2020understanding, fosco2020predicting}.  These models typically use Convolutional Neural Networks (CNNs) to extract features from images and predict pixel-level saliency. However, this pixel-level prediction is insufficient in two respects, when it comes to capturing how dynamic visual highlights drive visual attention in GUIs: firstly, pixel-level saliency cannot directly indicate how much attention a highlighted UI element has drawn, which is especially important when comparing highlighted elements with non-highlighted elements on GUIs; secondly, current models are designed for either constantly changing contexts like video streams~\cite{wang2018revisiting, Li2024} or static contexts like GUI screenshots~\cite{leiva2020understanding, jiang2023ueyes} or visualizations~\cite{wang2023scanpath}. They struggle to capture the temporal characteristics of attention in mixed GUI environments where static graphical elements are combined with dynamic content such as visual highlights, and dynamic changes in the underlying task.}

\revision{To address this gap, our study focuses on temporal element-level attention prediction: modeling how visual attention evolves in response to dynamic highlights over time. Achieving this requires not only predictive modeling but also a clear understanding of how dynamic visual highlights shape user behavior. We begin by analyzing the impact of visual highlights on user attention in a simulated drone monitoring task, where participants were asked to detect critical events with visual highlights indicating their occurrence. Gaze data collected during the task suggested that highlights lead to immediate and focused attention on the relevant elements. Based on this, we then aimed to predict user attention over time, both at the pixel and element levels. Since existing saliency models lack the capacity to process highlight dynamics explicitly, we developed the Highlight-Informed Saliency Model (HISM), which incorporates spatial and temporal features to predict normalized saliency (NS) for highlighted interface elements. Based on prior work and our observations, the following hypotheses guided our research:}

\revision{\textbf{Hypothesis 1 (H1).}  
Visual highlights will lead to a rapid increase in user attention to the highlighted element, improving detection performance but potentially reducing situation awareness (SA) due to less exploration of the overall interface.}

\revision{\textbf{Hypothesis 2 (H2).}  
Incorporating temporal features—such as the timing of the highlight—into the saliency model will enable more accurate predictions of when, how long, and how much attention is captured. Additionally, integrating task-specific state information will further enhance model performance, especially when highlights are not present.}


In sum, this paper makes the following contributions: \begin{itemize} \item We examine how visual highlights affect user attention and SA in monitoring interfaces, using NS metrics to quantify both the rapid attraction of attention and its temporal evolution. \item We propose HISM, a novel saliency model that integrates temporal and spatial features to predict dynamic attention shifts, outperforming state-of-the-art pixel-level models. \end{itemize}
Ultimately, by shedding light on the role of visual highlights and predictive modeling in GUI interaction, we hope to contribute towards the creation of more supportive, user-centric interfaces for monitoring tasks.
\section{LITERATURE REVIEW}


\subsection{Visual Highlighting in Monitoring Interfaces}


\revision{Monitoring tasks in dynamic interfaces such as multi-drone control panels require users to manage attention across multiple sources of information. As automation increases and system functionalities become more integrated, the interface itself can become visually dense and cognitively demanding, raising the risk of missed critical events~\cite{cheng2023novel,Wickens2021-vt}.  In such contexts, visual highlights—especially those using color for contrast—have proven effective in directing user attention to relevant elements~\cite{das2024shifting, feit2020}}.

Nevertheless, without careful design of the timing, duration, and frequency of such cues, the implementation of visual highlights can mask users' perception of critical information rather than facilitating effective task completion \cite{wan2024attention}. When multiple visual highlights appear in rapid succession or remain on screen for too long, they can compete for attention and inadvertently obscure other essential information. This challenge is particularly pronounced in multitasking scenarios, where users must divide their attention across multiple interface elements while managing simultaneous tasks \cite{wickens2007dual}. One well-documented instance of this problem is alarm floods, where multiple warning signals appear in close spatial and temporal proximity, overwhelming the user with dense and simultaneous notifications. As a result, users may struggle to prioritize information, leading to cognitive overload and loss of SA \cite{wan2019identifying}. 




\subsection{Navigating Monitoring Tasks: Situation Awareness and Gaze Behavior}


Maintaining SA is fundamental in monitoring tasks: it requires perceiving and comprehending environmental elements, then projecting future states to guide decisions and actions\cite{zhou2021using}. In multi-drone monitoring, for example, SA includes understanding real-time sensor data as well as potential hazards \cite{sun2022exploring,choi2022development}. SAGAT\cite{Endsley2000SAGAT} is often used to measure SA by freezing the task and querying the operator’s knowledge, though it traditionally targets air traffic scenarios. Accordingly, we refined it for multi-drone settings where participants must also address critical alerts. 

Evidence supports a strong correlation between SA scores and other physiological measures such as eye-tracking~\cite{zhang2020physiological}. Researchers found that frequent fixations on important events predict higher SA for those events~\cite{moore2010development}.  Furthermore, analysis of eye-tracking metrics reveals that higher fixation counts and longer dwell times on Areas of Interest (AOIs) containing information about environmental hazards correlate with higher SA. Zhang et al.~\cite{zhang2020physiological} provide a comprehensive review of how different eye-tracking metrics relate to SA, concluding that compared to involuntary physiological responses to environmental stimuli (e.g., blink rate and pupil dilation), conscious eye movements like fixations and saccades exhibit more significant correlations with SA.

\subsection{Predicting Visual Attention on GUIs: Temporal Saliency of Elements}

Among the various approaches for predicting users' visual attention, one of the most prominent ones is saliency prediction \cite{min2019tased, fosco2020predicting}. Such saliency prediction mostly produces pixel-level maps reflecting how likely each region of an image is to attract attention\cite{jiang2023ueyes}. \revision{However, in GUI contexts, pixel-level saliency may not accurately represent user attention on structured elements such as buttons, text, or icons. User attention is shaped not only by visual features like size, shape, and color, but also by the semantic relevance of these elements within the task \cite{gupta2018saliency,  wang2023scanpath}. When highlights are applied to specific elements, it is crucial to evaluate the entire targeted area and understand its
relative saliency compared to the overall interface.}

\revision{On the other hand, as human visual attention evolves over time \cite{traver2021glimpse}, there is a growing interest in predicting the saliency at different viewing times \cite{aydemir2023tempsal, wang2023scanpath}. While these approaches have advanced our understanding of temporal visual attention, they are not effective in capturing the influence of dynamic highlights in static GUIs. This mixed visual stimulus demands a new saliency model that can fully utilize both the temporal information, such as when the highlight occurs, and the spatial information in the GUIs. In response to this gap, our work proposes a new saliency model with two branches to integrate the temporal information of the visual cue together with the spatial information from the GUIs.} 

\section{Method}
\label{sec:method}

\subsection{Data Collection}
\label{sec:data_collection}

To investigate how visual highlights impact user attention and SA in a dynamic monitoring task, we conducted a controlled user study using a simulated multi-drone monitoring interface (see \autoref{fig:drone_interface}). Participants were tasked with continuously monitoring multiple drones, interpreting system alerts, and responding to critical situations in real-time. Throughout the experiment, eye-tracking data were recorded to analyze visual attention shifts and temporal gaze patterns in response to visual highlights.


\subsubsection{Monitoring Interface} 
\begin{figure}[h]
    \centering
    \includegraphics[width=0.8\columnwidth]{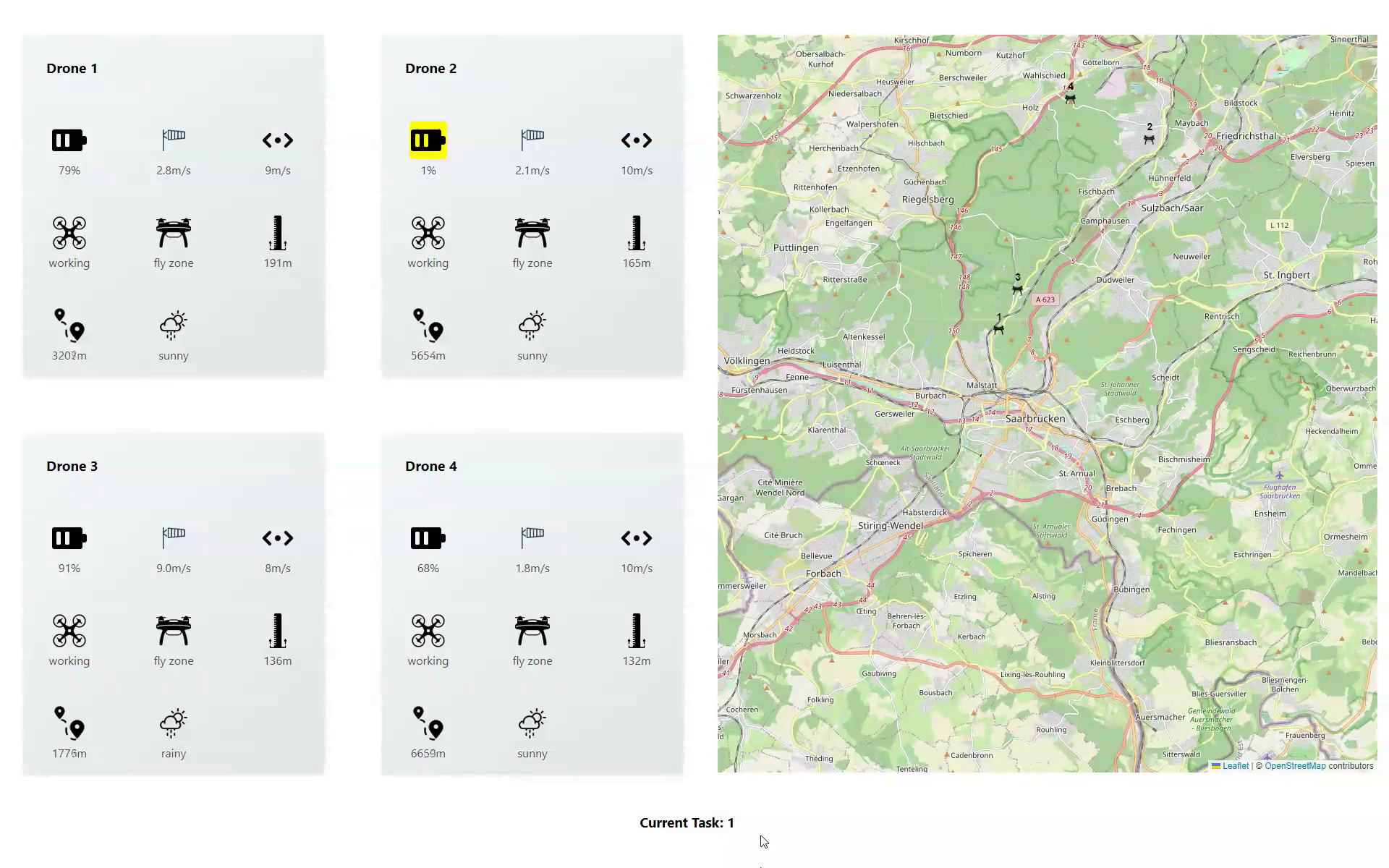}
    \caption{Drone monitoring interface highlighting a low-battery critical situation. 
    (A non-highlighted critical situation interface appears the same, but without the yellow mark.)}
    \label{fig:drone_interface}
\end{figure}

 Informed by previous research~\cite{Finkbeiner23,gregorio2021improving,sun2022exploring} and resembling existing multi-drone monitoring interfaces~\cite{djimavic3downloads,flytbasedronedelivery,sun2022exploring}, while being simple enough to be used in controlled lab studies and for collecting reliable gaze that, and easy to understand without prior experience in flying drones.  As illustrated in \autoref{fig:drone_interface}, the chosen interface combines icon-based elements, facilitating quick assimilation of drone parameters, with a map display to increase spatial awareness and task immersion. Each drone block features eight elements, showcasing core drone metrics along with a representative image (for simplicity, we refer to the combination of both the image and textual or numeric data as \emph{icon} in the following). These were chosen to cover different categories of data found to be relevant for drone monitoring~\cite{sun2022exploring}: 
\begin{enumerate}
    \item \textit{Safety Level and Alert} Battery level, wind speed, rotor condition, and no-fly-zone warnings determine the drone's current safety status. These values indicate potential critical situations.
    \item \textit{Telemetry Data}: Horizontal speed, altitude, and distance to the destination are data collected in real-time by sensors of the drone, providing insights into the drone's current operational mode and task progression. 
    \item \textit{Weather Data}: An icon about current weather conditions was incorporated to provide real-time environmental data, which might impact the drone's safety and task success.
\end{enumerate}
The icons visually symbolize these core metrics, which makes it easy to interpret the corresponding sensor values displayed below. Throughout the study tasks described below, the icons retained a static visual representation. Only the underlying values are updated according to the simulated state of the drone. This ensured visual consistency across the study conditions, which is important to isolate the effect of visual highlighting on users' gaze behavior.

As shown in \autoref{fig:drone_interface}, our design used a yellow background highlight to signal critical situations. After experimenting with various sizes and intensities, we determined that this struck the right balance between visibility and distraction prevention. We decided to focus on one established highlighting technique to thoroughly investigate the effect of visual cues on the users' task-driven attention during a monitoring task and to collect enough training data to develop deep learning-based visual saliency models that could predict this effect.


\subsubsection{Task Design}Each participant was tasked with completing four drone monitoring tasks, as indicated by a `current scene number' on the screen, each lasting 5 minutes. Participants were instructed to focus on two tasks: (1) identify critical situations and pressing the space bar to acknowledge their detection, and (2) at any time be aware of the state of all drones and correctly answer questions about icon values at random intervals. The design of each of these tasks is further described in the following.

\textit{Detecting Critical Situations:} Participants were asked to identify all critical situations by observing any changes in the indicator values for each drone block. Upon detecting a critical situation, they should press the space bar to acknowledge detection.
We separated each monitoring task into twenty 15-second intervals. These were initially randomized with respect to two aspects: 
(1)~the occurrence of a critical situation (80\% probability) and (2)~highlighting of the corresponding critical icon (50\% probability across critical situations).  As a result, during the study, participants encountered 64 critical situations (each lasting for 15s), where one of the four safety level icons (battery level, wind speed, rotor condition, and fly-zone warning) transitioned into a critical range, as defined in \autoref{tab:critical_situations}. In 33 of these cases, the corresponding icon was highlighted as described above. 

\begin{table}[t]
\centering
\renewcommand{\arraystretch}{1.5} 
\caption{Critical Situations and their Indicator Values}
\begin{tabular}{p{0.35\columnwidth} p{0.25\columnwidth} p{0.25\columnwidth}}
\hline
\textbf{Critical Situation} & \textbf{Indicator Value} & \textbf{Icon} \\
\hline
Low Battery & $<$10\% & \includegraphics[width=0.2in]{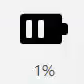} \\
\hline
Extreme Wind & $>$10 m/s & \includegraphics[width=0.2in]{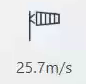} \\
\hline
Rotor Off & Off & \includegraphics[width=0.2in]{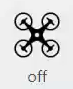} \\
\hline
No-fly Zone Warning & No-fly & \includegraphics[width=0.2in]{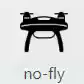} \\
\hline
\end{tabular}
\label{tab:critical_situations}
\end{table}

\begin{figure}[h]
\centering
\includegraphics[width=\columnwidth]{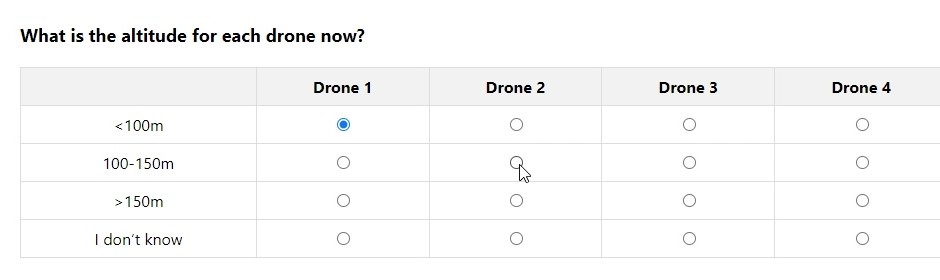}
\caption{Situation awareness question screenshot}
\label{fig:questions}
\end{figure}

\textit{Situation Awareness}: In addition to monitoring the drones for critical situations, participants were instructed to stay aware of the state of all icons across all drones, as well as the ones not related to critical situations (horizontal speed, altitude, distance to target, and weather condition). To assess participants' SA for these icons, our task design incorporated a SAGAT-based questionnaire whose design followed the recommendations provided by Endsley~\cite{endsley2000direct}. It appeared at randomized intervals, ranging from 30 to 60 seconds (but kept the same across participants). As a result, there was a total of 20 questionnaires during the whole experiment, with 2 appearing during a non-critical situation, 13 during a non-highlighted critical situation, and 5 during a highlighted critical situation.  Each questionnaire consisted of two questions, focusing on two icon values. The design shown in 
 \autoref{fig:questions} followed prior recommendations~\cite{endsley2000direct} to ensure that the SA assessment was short and as unintrusive to the monitoring task. 
 

\subsubsection{Procedure}

Each participant began the experiment with a detailed introductory video explaining the user interface and tasks. We used Tobii Pro Eye Tracker Manager’s five-point calibration. Additionally, we introduced four check pages—one per task—to confirm tracking accuracy and filter out poor-quality data (details in supplemental material). After calibration, participants completed a brief 90-second practice task (including all four critical situations and SA questions) to confirm their understanding of both the detection and question-answering procedures. They then undertook four tasks as detailed in the task design session.  Breaks were allotted between each of them, during which the eye-tracker was recalibrated to maintain tracking quality. Upon completing the last task, participants were asked to fill out a post-study questionnaire to gather their demographic data, feedback, and perceptions of the study (see supplementary material). 
At the end, participants were paid 15€ as compensation for their time. The study procedure and task were approved by the university's ethics committee.

%

\subsubsection{Participants}
\revision{We recruited a total of 28 participants (5 female, 23 male) from Saarland University, aged between 20 and 37 years (median age = 26). Most participants had no prior experience flying drones, with only five indicating intermediate to advanced experience. The sample size was determined based on two considerations. First, an a priori power analysis using G*Power indicated that at least 27 participants were required to detect a medium effect size ($d = 0.5$) with 80\% power at a 0.05 significance level using a two-tailed paired-sample $t$-test. Second, we performed a reliability check by repeatedly partitioning the participants into equal groups and computing the correlation between their fixation maps to ensure stability and representativeness of the gaze data. Further details of this validation procedure are provided in the supplementary material.}

\subsubsection{Hardware Setup}
\revision{The experiment was conducted using a 24-inch desktop monitor (52 × 32.5 cm, 1920 × 1200 px resolution), with participants seated at a fixed viewing distance of 60 cm. This setup yielded visual angles of approximately 45.4° horizontally and 30.5° vertically. Each pixel corresponded to roughly 0.0271 cm, making the 25-pixel dispersion threshold used for fixation detection equivalent to a visual angle of approximately 0.65°. To capture high-fidelity gaze data, we used a Tobii Pro Fusion eye tracker operating at 250 Hz, positioned below the screen and angled upward to optimize accuracy. Calibration was performed using the Tobii Pro Eye Tracker Manager.}

\subsection{Data Analysis}

\subsubsection{Data filtering and preprocessing}
\label{sec:data_processing}
Each gaze point was captured as an (x,y) screen coordinate, together with its corresponding timestamp. From the raw gaze points, fixations were identified based on specific criteria: low dispersion (25\, px) and adequate duration (50\, ms), using the fixation detection function from the PyGaze package~\cite{dalmaijer2014pygaze}. We ensured data quality in two steps: first, we verified dataset comprehensiveness using correlation-coefficient checks; second, we excluded intervals with poor accuracy in gaze offsets. Full details appear in the supplementary materials.




For the subsequent analysis, gaze data were segmented into 15-second task intervals corresponding to the duration of potential critical situations, as described in \autoref{sec:data_collection}. Additionally, the timestamps of participants' button presses, which indicated the detection of critical situations, were aligned with the gaze data timestamps. The dataset was further complemented with participants' responses to the SA questions.

\subsubsection{Gaze and Performance Metrics}
To assess the impact of visual highlighting on detection performance and SA, we analyze gaze behavior through two primary categories: engagement and exploration. Engagement metrics capture how users direct attention toward specific elements, while exploration metrics describe how users scan the interface and shift focus across different areas.

Engagement is measured by tracking \textit{fixation count}, \textit{fixation duration}, and \textit{revisits}, which provide insights into how users allocate attention to critical areas such as icons indicating system alerts or SA-related queries. In contrast, exploration behavior is characterized by \textit{mean saccade amplitude}, \textit{scanpath length per second}, and \textit{AOI transition rate}, which reflect broader search patterns and visual scanning strategies across the interface.

To further analyze visual attention distribution, we generated continuous saliency maps by synchronizing the collected gaze data with the UI's frame rate and extracting fixation details as specified in \autoref{sec:data_processing}. Fixation points from all participants were aggregated and processed using a Gaussian kernel, a commonly used technique in prior research, before normalizing the resultant saliency map within a [0,1] range.


\subsubsection{Statistical Analysis}

The analysis begins with a static approach, where gaze and performance metrics are aggregated over the entire 15-second task interval to assess the overall effects of highlighting. This is followed by a temporal approach, which examines how the influence of highlights evolves over time, providing deeper insights into the dynamic interaction between visual cues and user behavior.

For each analysis, statistical hypothesis testing was conducted after assessing normality using the Shapiro-Wilk test. For normally distributed continuous variables, we applied an independent two-sample $t$-test, reporting the test statistic $t$, degrees of freedom ($df$), and p-value ($p$). For non-normal or non-continuous variables, we used the Mann-Whitney $U$ test (Wilcoxon rank-sum test), reporting the test statistic $U$, sample sizes $n_1$ and $n_2$, and the p-value $p$.We considered $p<0.05$ as significant. All analyses were conducted at the participant level, with each data point representing the average of all trials per participant.

\section{Empirical Results}


\subsection{Influence of Highlighting on Detecting Critical Situations} 
\label{sec:highlight_detection_verification}

\begin{table}[h]
\caption{Summary of Detection Performance (Mean and SD)}
\centering
\small 
\renewcommand{\arraystretch}{1.1} 
\begin{tabular}{p{2.2cm}p{2cm}p{2cm}} 
\hline
\textbf{Detection Result} & \textbf{Critical} & \textbf{Non-Critical} \\
\hline
Identified as critical (\%) & 
$85.68 (1.77)$ & 
$2.71 (0.30)$ \\
Identified as non-critical (\%) & 
$14.32 (1.77)$ & 
$97.29 (0.30)$ \\
\hline
\end{tabular}
\label{tab:DetectionPerformance}
\end{table}

\autoref{tab:DetectionPerformance} presents an overview of detection performance across participants. \revision{Specifically, a trial was labeled “non-critical” if the participant did not press the space bar before the end of the 15-second interval.}. Most critical situations were correctly identified (85.68\%), while false alarms were minimal (2.71\%). Given the negligible rates of false alarms, in the following, we focus only on the analysis of correctly detected critical situations. 

\begin{table}[h]
\caption{Hit Rates and Response Time in Different Highlight Conditions(Mean and SD)}
\centering
\small
\setlength{\tabcolsep}{4pt}
\renewcommand{\arraystretch}{1.2}
\begin{tabular}{p{2.5cm}cc}
\hline
\textbf{Metric}                      & \textbf{Highlight}       & \textbf{No Highlight} \\
\hline
Hit Rate (\%)               & $96.18(0.40)$         & $74.76(0.84)$      \\
Response Time (s)          & $1.33(0.31)$          & $5.16(0.79)$       \\
\hline
\end{tabular}
\label{tab:Hits_Response}
\end{table}



As indicated in the first row of the \autoref{tab:Hits_Response}, participants detected significantly more critical situations when highlights were present (\textit{U(28,28)} = 819.0, $p < .001$). Additionally, the second row of \autoref{tab:Hits_Response} shows that response times were significantly shorter with highlights, confirming that visual cues facilitate faster detection (\textit{ U(28,28)} = 0.0, $p < .001$). Together, these results confirm that visual highlighting enhances both accuracy and speed in detecting critical situations, aligning with results from previous findings~\cite{das2024shifting,feit2020detecting}.

\begin{figure}[h]
\centering
\includegraphics[width=0.8\columnwidth]{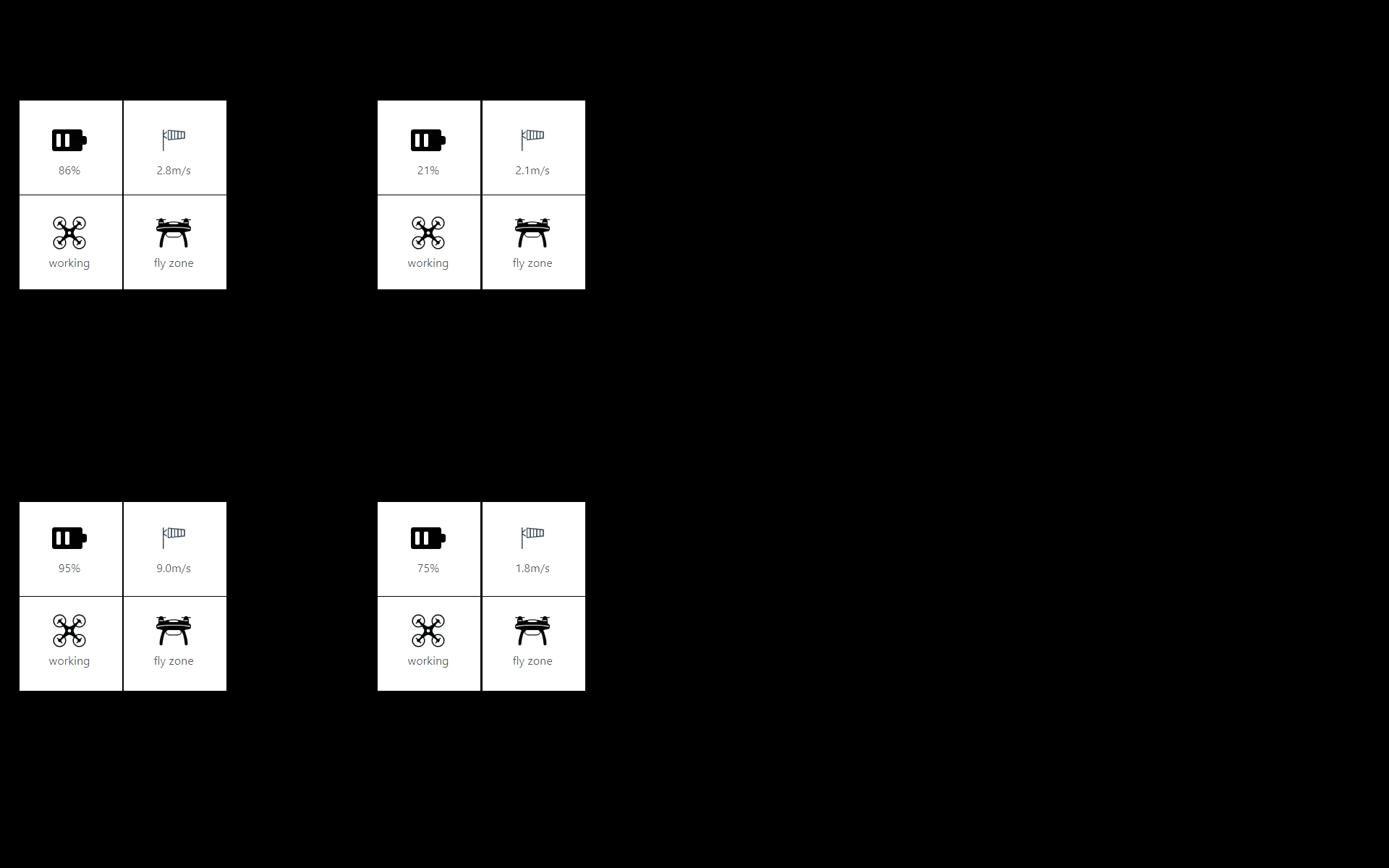}
\caption{AOIs related to the critical situation.}
\label{fig:AOI}
\end{figure}

To better understand how highlights drive these improvements, we analyzed the gaze of the participants when engaging with the targeted icon in the case of a critical situation. Each AOI, as illustrated in \autoref{fig:AOI}, encompassed the icon and its associated text, spanning $142\mathrm{px} \times 128\mathrm{px}$. We calculated the three engagement metrics -- fixation count, fixation duration, and revisits -- on the AOI related to a critical situation for those task intervals where participants successfully detected the critical situation.

\begin{table}[h]
\centering
\small 
\setlength{\tabcolsep}{2pt} 
\renewcommand{\arraystretch}{1.1} 
\caption{Verification Gaze Metrics in Highlight and No-Highlight Conditions (Mean and SD)}
\label{tab:AOIGazeMetricsComparison}
\begin{tabular}{lcc}
\toprule
\textbf{Gaze Metric} & \textbf{Highlight} & \textbf{No Highlight} \\
\midrule
Fixation Count         & 6.37(3.76)   & 8.69(4.79)   \\
Fixation Duration (s)  & 5.36(1.58)   & 8.02(2.58)   \\
Revisits               & 2.18(0.52)   & 2.73(0.62)   \\
\bottomrule
\end{tabular}
\end{table}

\autoref{tab:AOIGazeMetricsComparison} shows that there were fewer fixations in highlighted trials compared to non-highlighted trials (\textit{U(28,28)} = 519.0, $p < .01$). Fixation duration was also shorter with highlight ($t(27) = 4.58$, $p < .001$). Additionally, there were fewer revisits when highlights were present (\textit{U(28,28)} = 580.5, $p < .001$).

These results suggest that visual highlights enable participants to notice critical situations more quickly, reducing the time required to process the information. However, they also lead to shorter fixation durations and fewer revisits, indicating that users may engage with the highlighted information less thoroughly. 

\subsection{Influence of Highlighting on Situation Awareness} 
\label{sec:highlight_sa_search}

\begin{table}[h]
\centering
\footnotesize
\setlength{\tabcolsep}{4pt} 
\renewcommand{\arraystretch}{1.2} 
\caption{Situation Awareness and Gaze Metrics under Different Highlight Conditions (Mean and SD)}
\label{tab:SA_h_nh}
\begin{tabular}{lcc}
\toprule
\textbf{Metric} & \textbf{Highlight} & \textbf{No Highlight} \\
\midrule
SA Score & 0.49 (0.20) & 0.51 (0.22) \\
\midrule
Fixation Count & 1.27 (1.32) & 2.62 (1.96) \\
Fixation Duration (s) & 0.10 (0.06) & 0.19 (0.11) \\
Revisits & 0.11 (0.27) & 0.70 (0.36) \\
\midrule
Mean Saccade Amplitude (px) & 98.97 (37.33) & 109.66 (41.36) \\
Scanpath Length per Second (px/s) & 892.64 (336.73) & 1028.71 (390.19) \\
AOI Transition Rate (/s) & 2.54 (1.06) & 3.08 (1.22) \\
\bottomrule
\end{tabular}
\end{table}








Similarly, we analyze the SA and the relevant gaze behavior across the entire 15s task trial when critical situations happen. Our analysis revealed that SA scores were slightly lower in the highlight condition, compared to the no-highlight condition, as shown in the \autoref{tab:SA_h_nh}, although this difference was not statistically significant. However, we found significantly different gaze behavior when calculating the gaze behavior metrics critical for maintaining SA. The participants' engagement with the SA-queried icons was significantly higher in the no-highlight condition. Specifically, fixation count, fixation duration, and revisits were significantly higher in the no-highlight condition (\textit{U(28,28)} = 697.0, $p < .001$; $t(27) = 4.11$, $p < .001$; \textit{U(28,28)} = 723.5, $p < .001$, respectively).

\revision{Additionally, exploration behaviors—characterized by broader search patterns—were reflected in significantly higher scanpath length per second (\textit{U(28,28)} = 607.0, $p < .05$) and AOI transition rate (\textit{U(28,28)} = 616.0, $p < .05$) in the no-highlight condition. These increases indicate more extensive visual scanning across the interface. However, mean saccade amplitude, another indicator of search breadth, did not show a significant difference between conditions (\textit{U(28,28)} = 525.0, $p = 0.271$).}

These findings indicate that while highlights accelerate the detection of critical icons, they do not necessarily increase engagement with other elements. Reduced fixations, durations, and revisits point to a potential narrowing of attention at the expense of broader SA. To explore this trade-off over time, we analyze saliency maps to see whether highlights broaden or tunnel user attention across the interface.

\subsection{Temporal Visual Attention Analysis} 
\label{sec:temporal_analysis}

\begin{figure*}[t]
\centering

\begin{minipage}{.05\textwidth}
\raggedright
\textbf{H}
\end{minipage}%
\begin{minipage}{.95\textwidth}
  \begin{subfigure}{.2\textwidth}
    \includegraphics[width=\linewidth]{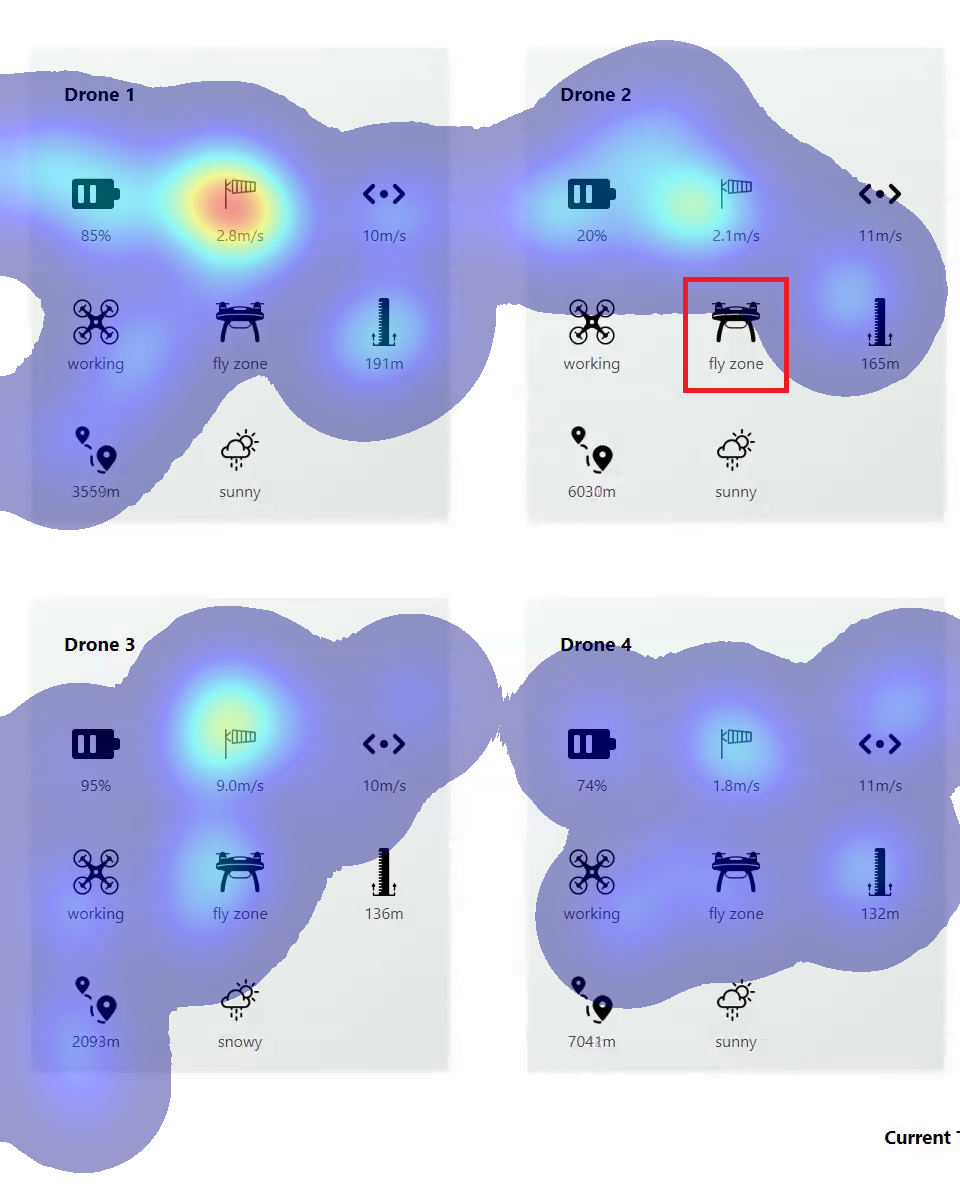}
  \end{subfigure}
  \hspace{0.5em}
  \begin{subfigure}{.2\textwidth}
    \includegraphics[width=\linewidth]{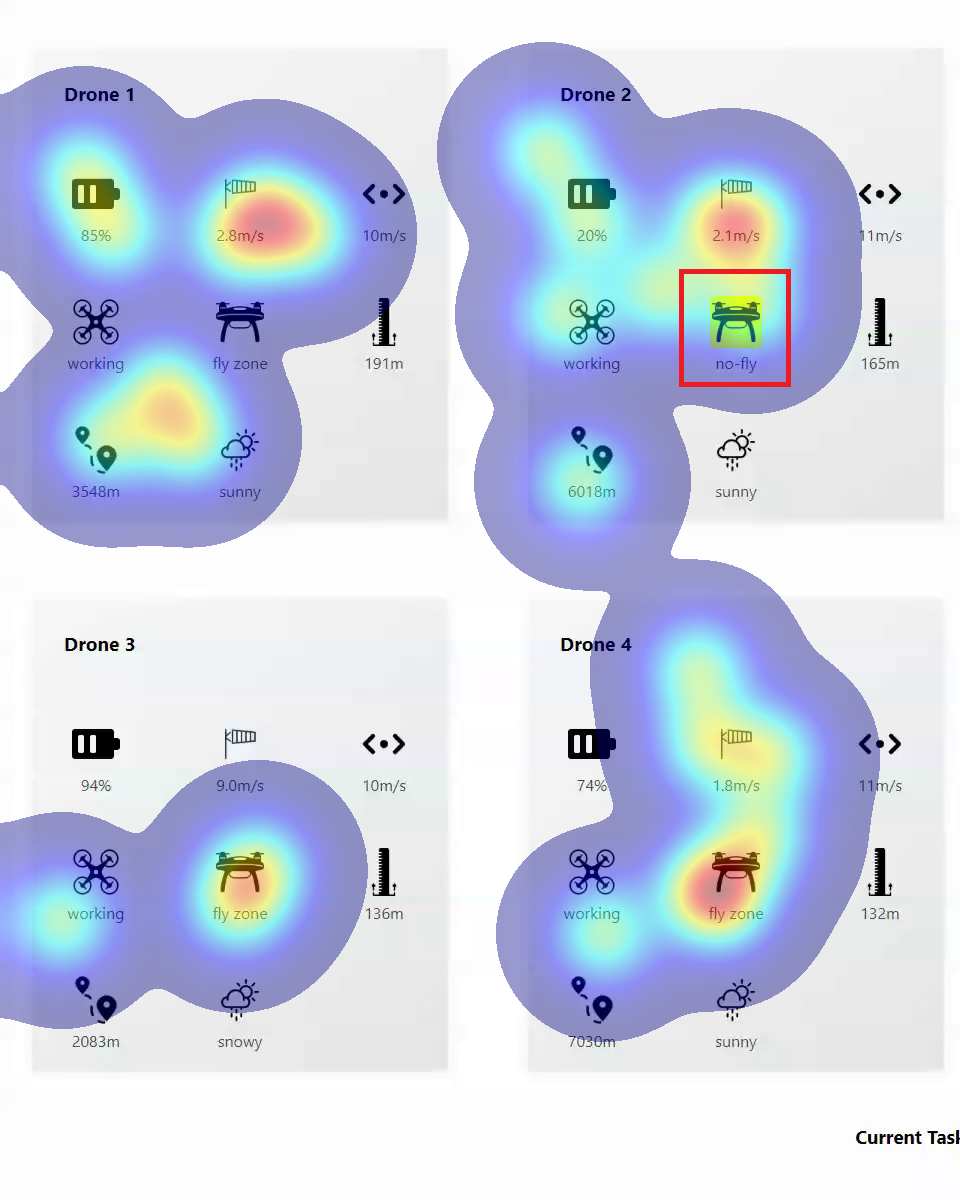}
  \end{subfigure}
  \hspace{0.5em}
  \begin{subfigure}{.2\textwidth}
    \includegraphics[width=\linewidth]{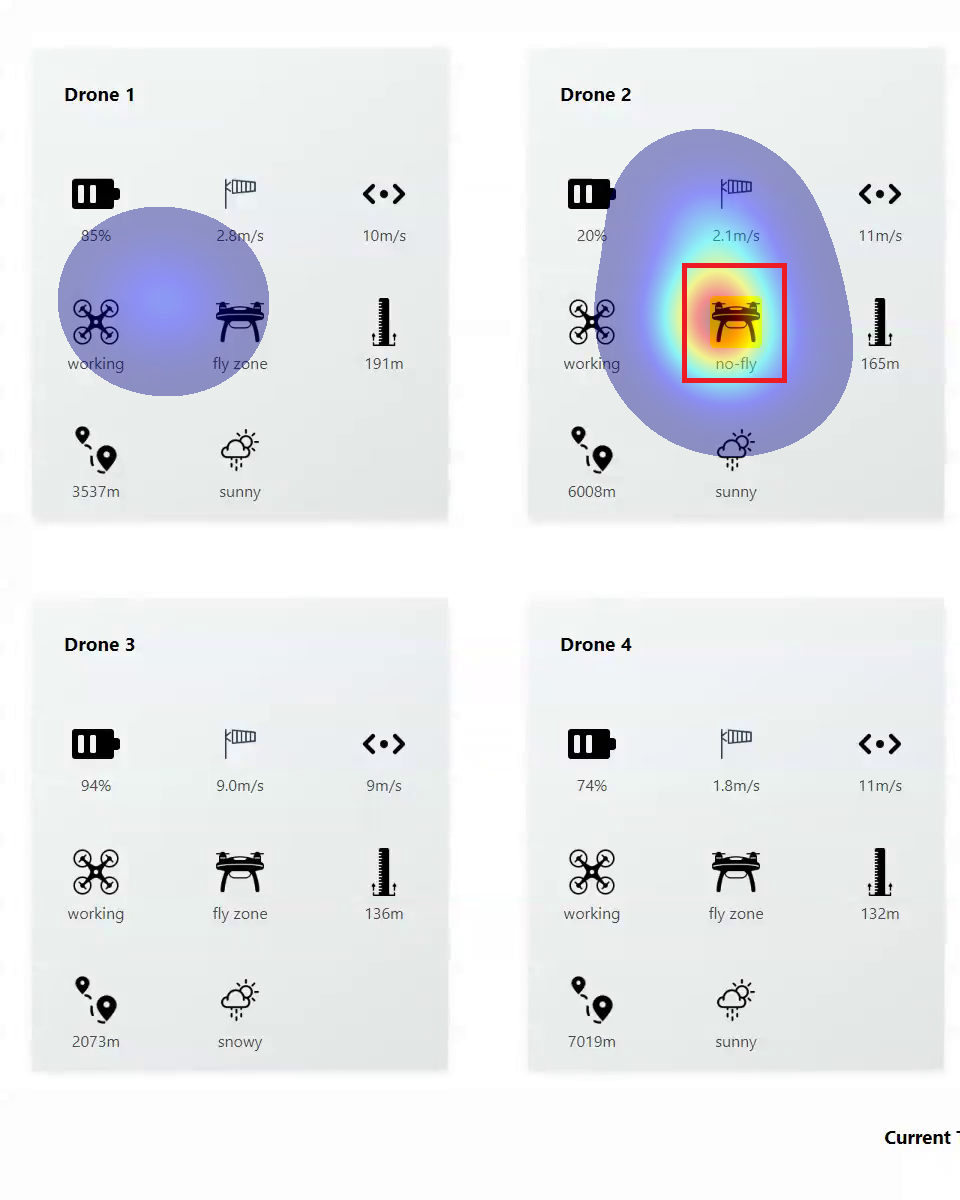}
  \end{subfigure}
  \hspace{0.5em}
  \begin{subfigure}{.2\textwidth}
    \includegraphics[width=\linewidth]{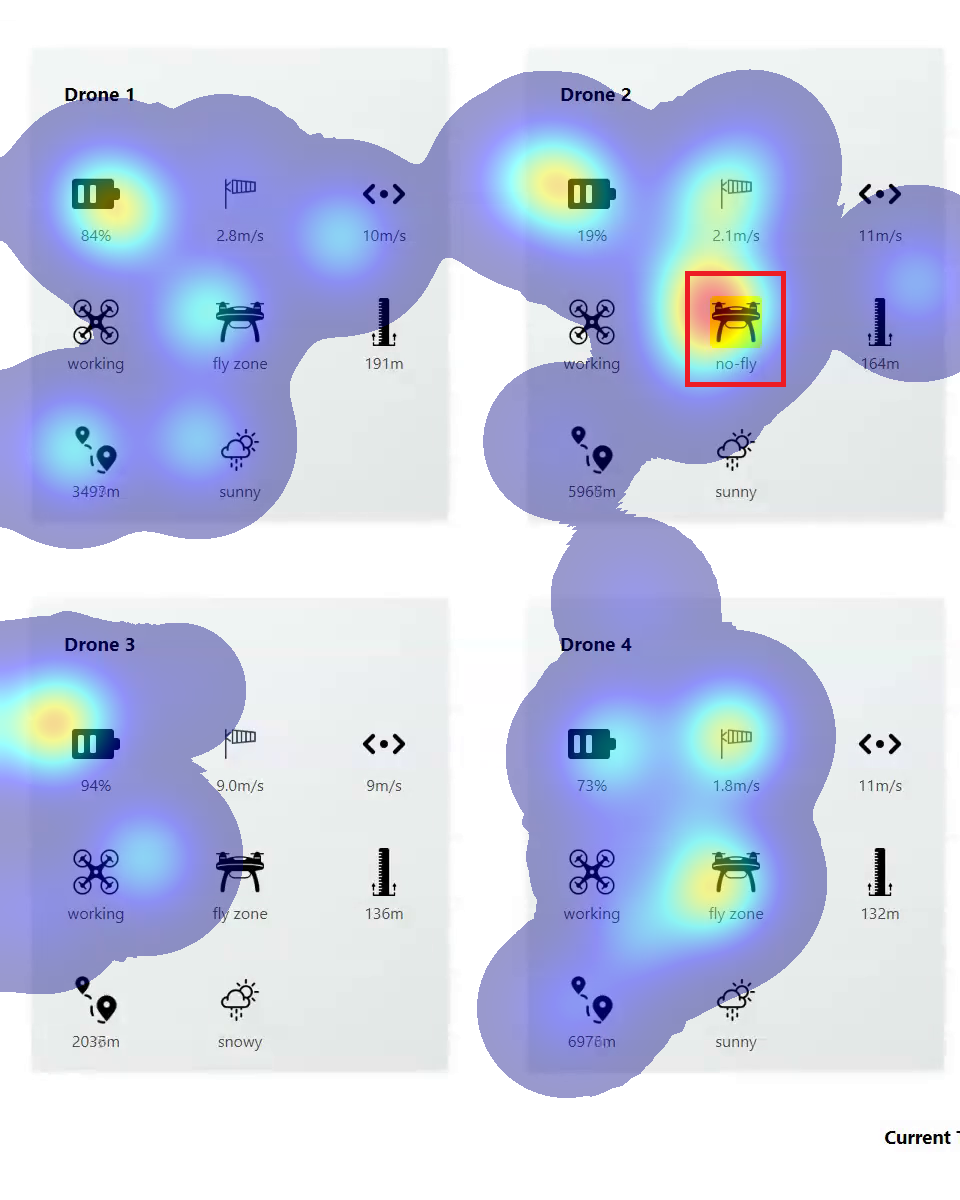}
  \end{subfigure}
\end{minipage}

\vspace{0.5em}

\begin{minipage}{.05\textwidth}
\raggedright
\textbf{NH}
\end{minipage}%
\begin{minipage}{.95\textwidth}
  \begin{subfigure}{.2\textwidth}
    \includegraphics[width=\linewidth]{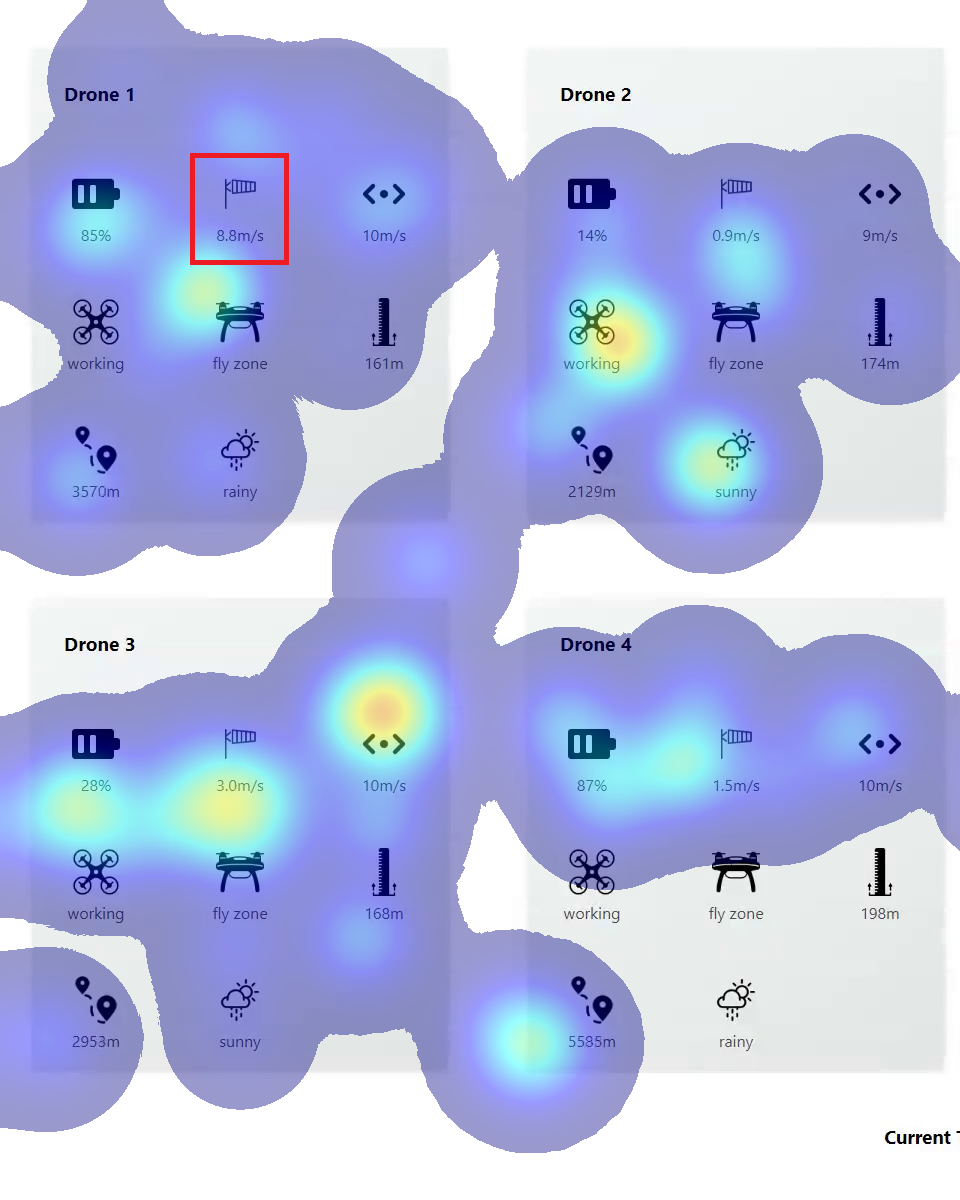}
    \caption*{T0 (1s Before CS)}
  \end{subfigure}
  \hspace{0.5em}
  \begin{subfigure}{.2\textwidth}
    \includegraphics[width=\linewidth]{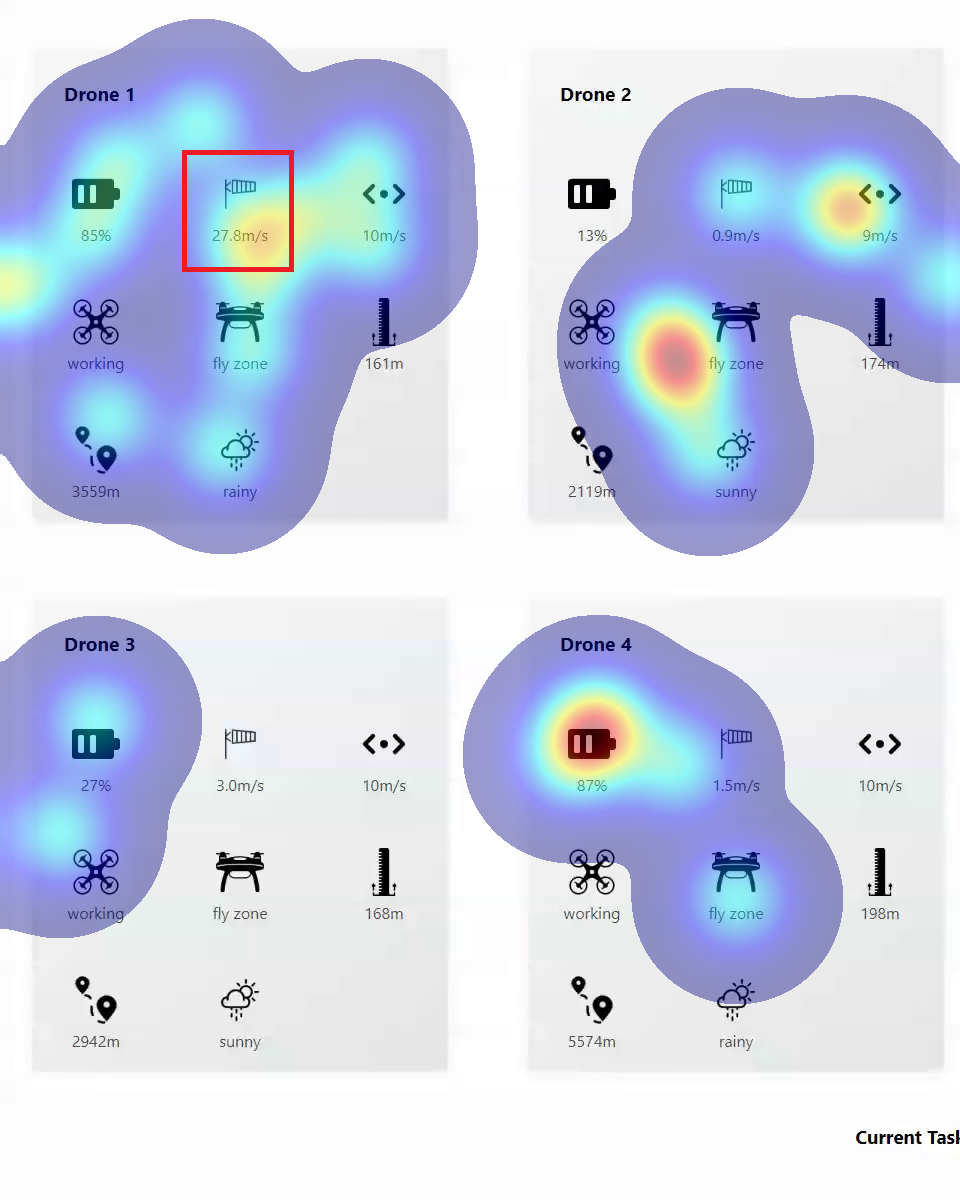}
    \caption*{T1 (CS Happened)}
  \end{subfigure}
  \hspace{0.5em}
  \begin{subfigure}{.2\textwidth}
    \includegraphics[width=\linewidth]{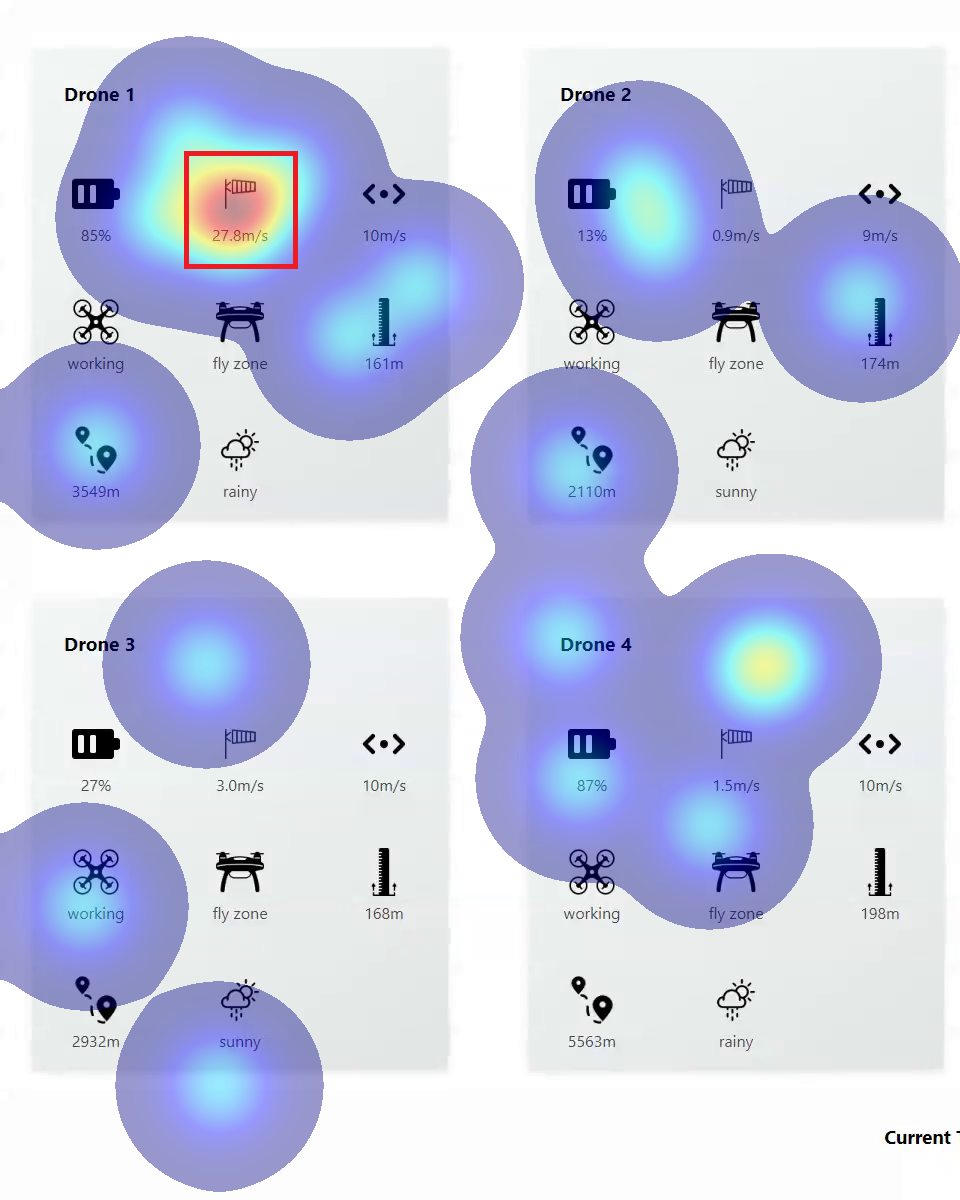}
    \caption*{T2 (1s After CS)}
  \end{subfigure}
  \hspace{0.5em}
  \begin{subfigure}{.2\textwidth}
    \includegraphics[width=\linewidth]{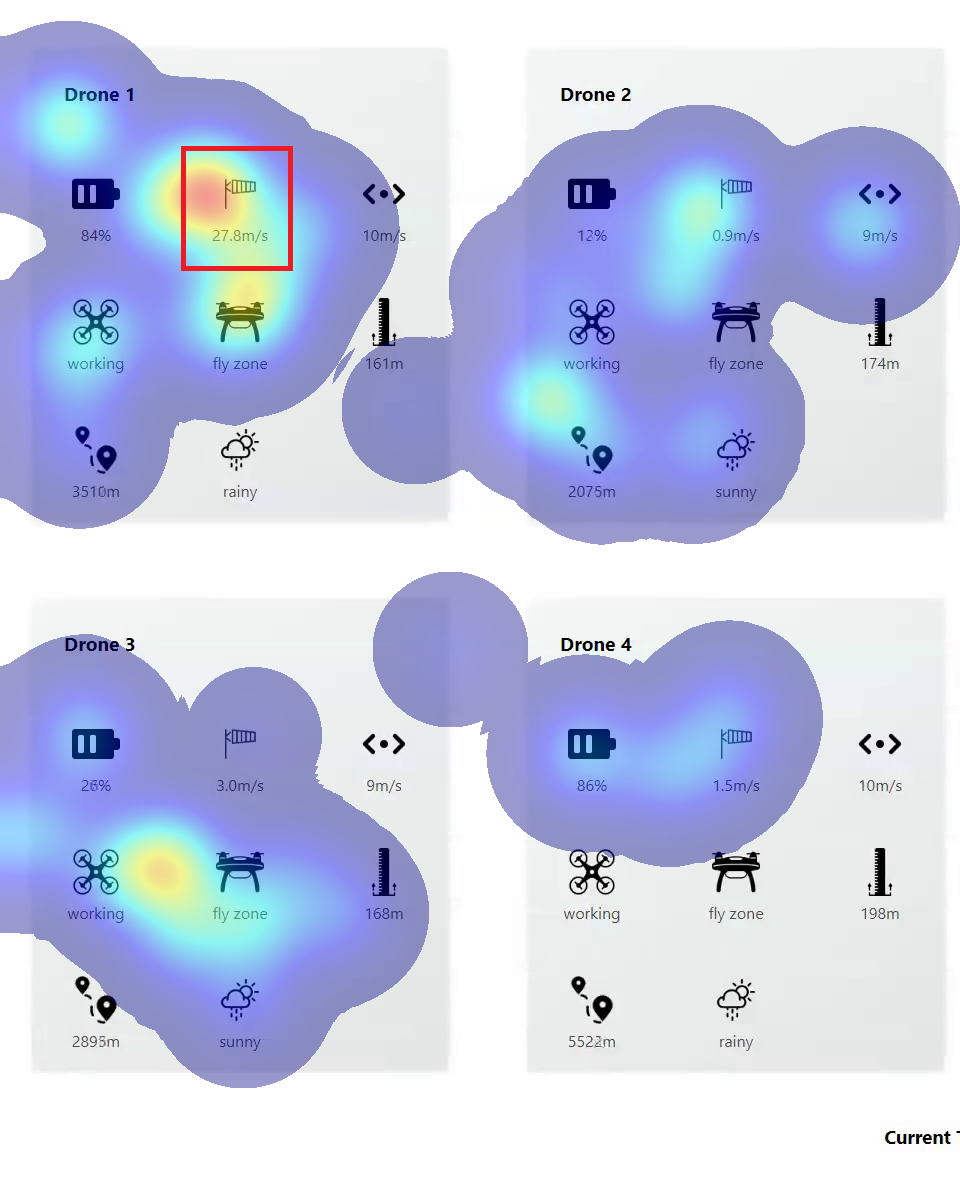}
    \caption*{T3 (5s After CS)}
  \end{subfigure}
\end{minipage}

\caption{Shifts in attention during a critical situation (CS) with (Top) and without (Bottom) highlights. Icons indicating a critical situation are marked in red (not visible during the study).}
\label{fig:visual_attention_temp}
\end{figure*}

To analyze the participants' temporal visual attention changes, we generated saliency maps for the interface frames and visualized them at specific timestamps. As shown at the top of \autoref{fig:visual_attention_temp}, the saliency maps reveal a clear tunneling effect caused by the visual highlight, which directs nearly all user attention to the highlighted icon shortly after its appearance. This effect is particularly evident in the initial moments (T1 and T2), where the highlight effectively draws attention to the critical situation icon. However, this focused attention comes at the cost of reduced engagement with other parts of the interface. As the participant identifies the critical situation, their attention begins to shift back toward the broader interface, gradually recovering the perception of the global picture. In contrast, in the no-highlight condition (bottom of \autoref{fig:visual_attention_temp}), this effect is less pronounced. Without the visual cue, attention is initially more distributed, and participants take longer to focus on the critical situation. As a result, their gaze behavior remains more varied over time, suggesting a broader search strategy that may help maintain awareness of other elements on the interface. This contrast further underscores the trade-off of visual highlights: while they facilitate rapid detection, they may temporarily suppress exploration and reduce engagement with non-highlighted areas.


Building on these observations, we analyzed the normalized saliency (NS) of a critical icon over time \cite{gupta2018saliency,wang2023scanpath}. 
The NS of a visual element is defined as the saliency of that element relative to the saliency of all elements within the GUI. The element saliency can be computed by integrating the pixel-level saliency density over the area covered by an element. Formally, we define the normalized saliency ($NS$) of an element $e_k$ at time slice $t_i$ as below:

\begin{equation}\label{eq:normalized_saliency}
NS(e_{k}, t_{i}) = \frac{S(e_{k}, t_{i})}{\sum_{j=1}^{n} S(e_{j}, t_{i})}
\end{equation}
where $S(e_k, t_i)$ is the saliency of element $e_k$ at timestamp $t_i$. This allows us to track the saliency of the potentially highlighted element ($e_h$) over the entire temporal trajectory $T = \{t_1, ..., t_m\}$.

Having defined NS as a measure of relative attention distribution, we now use it to quantify the saliency dynamics of the critical situation icon against all other icons over time. Specifically, we compute NS at 0.1-second intervals, enabling precise, time-resolved analysis of how attention dynamically shifts in response to the critical situation. 

\begin{figure*}[t]
\centering
\begin{subfigure}{0.48\textwidth}
  \includegraphics[width=\linewidth]{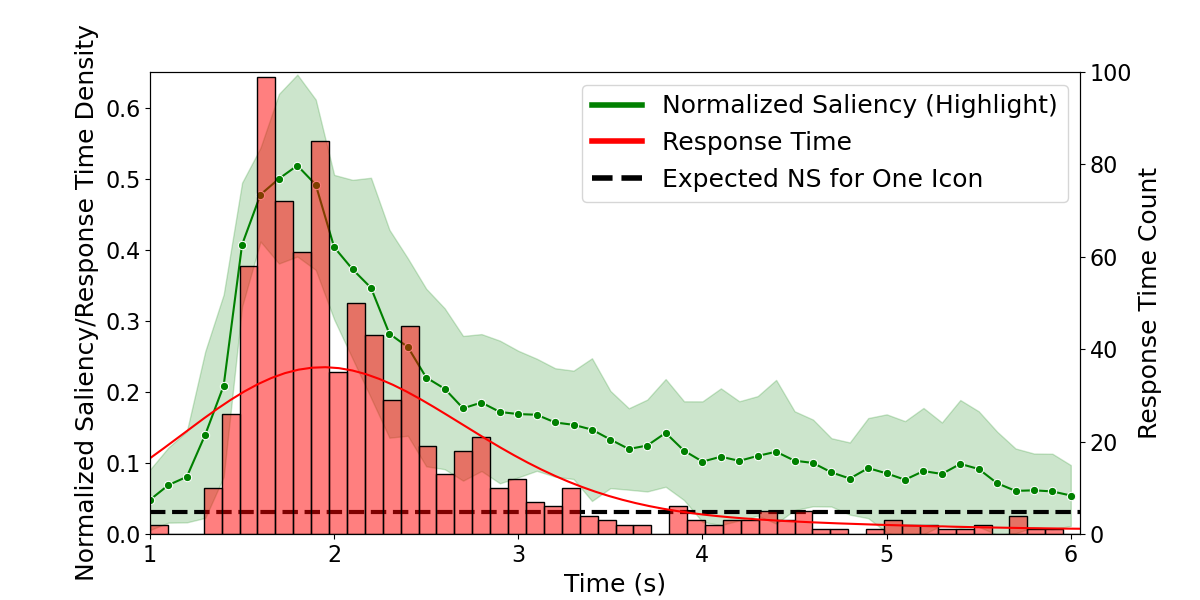}
  \caption{Highlight}
  \label{fig:ns_rt_h}
\end{subfigure}
\hfill
\begin{subfigure}{0.48\textwidth}
  \includegraphics[width=\linewidth]{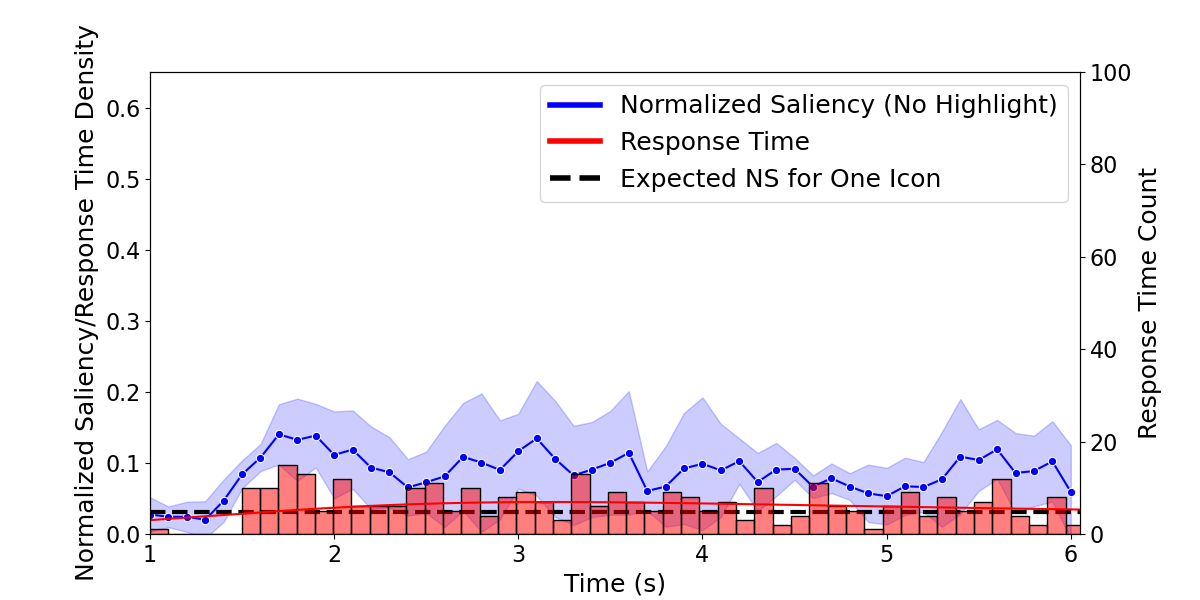}
  \caption{No Highlight}
  \label{fig:ns_rt_nh}
\end{subfigure}
\caption{Comparison of NS Changes Over Time (a) with Highlight. (b) without Highlight.}
\label{fig:ns_comparison}
\end{figure*}

\autoref{fig:ns_rt_h} illustrates the NS trends for the icon with highlight during the first five seconds following the onset of a critical situation. We observe a rapid surge in NS within the first second, peaking at values around 0.5. This sharp increase corresponds with the intense visual attention drawn by the highlight, as depicted in the heatmap at T1 in Figure \autoref{fig:visual_attention_temp}. However, this peak in visual attention is only temporary; the NS begins to decline shortly after reaching the peak, specifically after 0.6 to 1 second. This pattern of NS change aligns closely with the distribution of response times across these intervals, which also shows a rapid increase followed by a decrease. The average response time (M = 1.33 seconds) closely follows the NS peak, indicating that NS is an effective indicator of the timing at which users respond to highlighted information.

By contrast, in the no-highlight condition, NS remains low and stable throughout the first five seconds, as shown in blue in \autoref{fig:ns_rt_nh}. Unlike the sharp peak observed with highlights, NS only slightly increases to 0.1 following the onset of the critical situation, indicating a more gradual detection process. Across the five-second window, participants had a consistent probability of detecting the situation at any given timestamp, without a dominant moment of focus. Notably, by the final 0.5 seconds, NS values in the highlight condition (\autoref{fig:ns_rt_h}) rapidly diminish, approaching those in the no-highlight condition (\autoref{fig:ns_rt_nh}). This suggests that after the initial highlight-driven attention spike, users gradually redistribute their gaze, bringing NS closer to the expected random distribution level of 1/32.

\begin{figure}[h]
\centering
\includegraphics[width=\columnwidth]{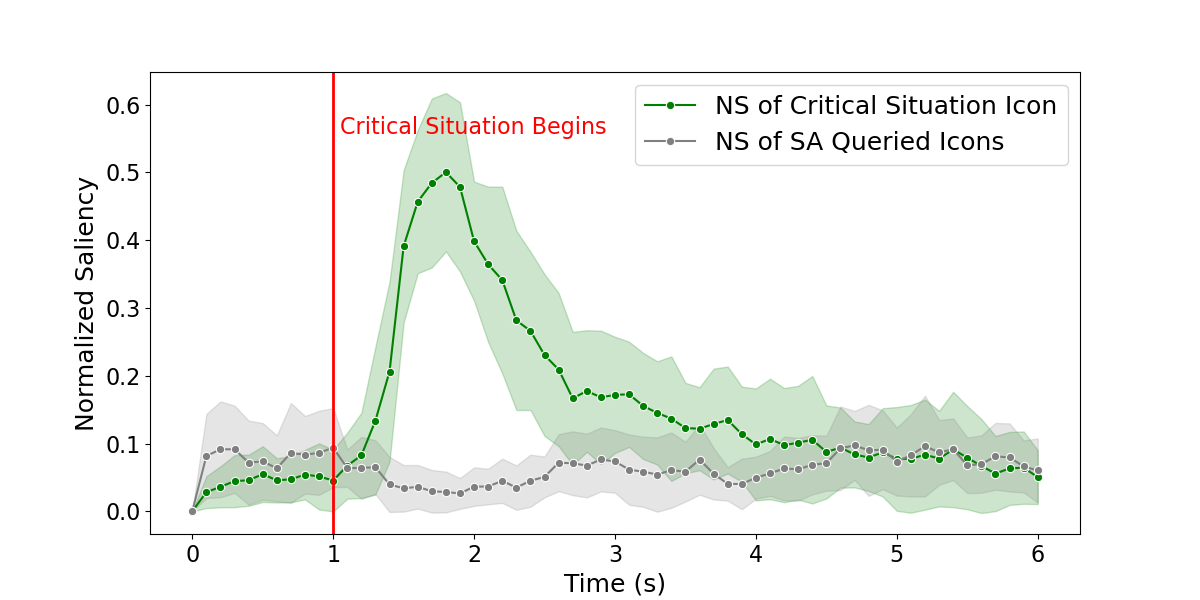}
\caption{Comparative NS changes for the critical situation icon and the SA queried icons, highlighting the shift in user attention.}
\label{fig:ns_cs_sa}
\end{figure}

We calculated the NS for the SA queried icons over a six-second range, including one second before the critical situation began. As shown in \autoref{fig:ns_cs_sa}, the NS of the critical situation icon (green) is already slightly higher than that of the SA queried icons (gray) in the 0 to 1 second range, indicating that participants allocated more attention to the SA-related icons when no highlight was present. Following the onset of the critical situation (1 to 6 seconds), there is a clear shift in attention from the SA icons to the highlighted icon. This is evident as the NS of the SA icons decreases while the NS of the critical situation icon surges, before gradually returning to its previous level as the NS of the critical situation icon declines. Statistical analysis confirmed a significant negative correlation between the NS values of the critical situation and SA-queried icons ($r = - 0.24$, $p<.001$), reinforcing the observed shift in visual attention.


 We conclude that the NS metric is a robust and precise indicator of user attention on specific targets relative to other AOIs. \revision{Compared to gaze metrics such as AOI statistics or frame-level saliency maps, the NS metric captures the relative visual saliency of the highlighted element in comparison to all other elements on the interface. It provides a single, interpretable value that reflects the proportion of user attention allocated to the highlight relative to the rest of the interface.} It measures how quickly, how much, and for how long visual highlights attract attention, and reveals how these shifts in attention influence detection responses and overall SA, making it a more effective tool for temporal visual attention analysis.

\section{Predicting Temporal Attention with Dynamic Highlighting}
\label{sec:prediction}

In this section, we explore whether we are able to predict these empirical observations regarding the impact of visual highlights on the temporal visual attention of users. \revision{We began by applying existing saliency models, including the bottom-up ITTI saliency map model \cite{itti2001computational}, and data-driven approaches \cite{reddy2020tidying,wang2018revisiting}, to estimate pixel-level attention.} We then shifted to a temporal, element-level prediction task, aiming to directly model the normalized saliency (NS) of the highlighted icon over time (as described in \ref{sec:temporal_analysis}). \revision{To support this, we introduced two key input representations to capture dynamic attention changes arising from both bottom-up and task-driven processes: a highlight vector, indicating when the highlight is active or absent across time steps, and a task vector, encoding the drone state values of the targeted icon. By combining these inputs with temporally-aware model architectures such as LSTMs and transformer encoders, our model significantly outperforms state-of-the-art baselines, enabling accurate predictions of attention dynamics in critical monitoring tasks.}

\subsection{Temporal Pixel-Level Saliency Prediction}
\label{subsec:pixel-level prediction}
\subsubsection{Saliency Dataset Generation}

For the first part of the temporal visual attention prediction, we focused on how users' gaze behavior changes during the transition from a non-critical to a critical situation. We analyzed gaze data from one second before to five seconds after the onset of each critical situation to capture visual attention dynamics and the influence of visual highlights. Specifically, we synchronized the gaze data with the UI's frame rate (1/24 per second) and extracted fixation details as described in Section \ref{sec:data_processing}. Following established methods~\cite{jiang2023ueyes,wang2018revisiting}, continuous saliency maps were created by merging fixation points from all participants during 42ms and applying a Gaussian kernel with a window size of 35px to smooth them \cite{bylinskii2018different}.
We thus obtained 4,608 smoothed saliency maps (of which 60\% were used for training, 10\% for validation, and the remaining 30\%  for testing), with each map corresponding to a specific frame of the monitoring interface. The paired saliency maps and corresponding UI frames formed the input and output data used for training and evaluating our models.

\subsubsection{Evaluation Metrics}

We assessed the performance of pixel-level saliency models with four commonly used metrics~\cite{bylinskii2018different}:  \textit{AUC (Area Under the ROC Curve)} measures the model's ability to classify fixation points across varying thresholds, calculating true and false positive rates across thresholds; \textit{NSS (Normalized Scanpath Saliency)} calculates the mean normalized saliency at fixation points, penalizing false positives; \textit{SIM (Similarity)} measures the intersection of two normalized saliency distributions as histograms, with a score of one indicating perfect overlap and zero indicating no overlap, and is sensitive to missing values and Gaussian blur, favoring partial matches; and \textit{CC (Pearson's Correlation Coefficient)} assesses the linear correlation between predicted and ground-truth saliency maps, symmetrically penalizing false positives and negatives and remaining resilient to linear transformations. In the development of our model, we also incorporated the Kullback-Leibler divergence (KL) into the loss function to evaluate the discrepancy between the predicted saliency and ground truth distributions. 

\subsubsection{Model Choice and Implementation}

\revision{We evaluated three models: ITTI \cite{itti2001computational}, a bottom-up model using low-level visual features; SimpleNet ~\cite{reddy2020tidying}, an encoder–decoder with ResNet backbone; and TASED-Net ~\cite{min2019tased}, which integrates spatiotemporal features using 3D convolutions.}

\revision{For the training process of the two data-driven saliency models, while SimpleNet was originally trained using KL\cite{shin2022scanner} and CC \cite{reddy2020tidying}, we extended the loss function to include SIM, resulting in the combined loss: 10KL – 3CC – 2SIM. This formulation improved alignment with ground truth and reduced false positives. Since TASED-Net requires sequences of frames (we used 32 per batch), its application was limited by the dataset size. To address this, we froze the encoder and fine-tuned only the decoder. We also modified its original loss to KL – 0.5CC – 0.1SIM, following prior work \cite{chang2021temporal}, which improved predictive accuracy and robustness in our task.}

\begin{figure*}[t]
\centering

\begin{minipage}{.05\textwidth}
\raggedright
\textbf{IT}
\end{minipage}%
\begin{minipage}{.95\textwidth}
  \begin{subfigure}{.2\textwidth}
    \includegraphics[width=\linewidth]{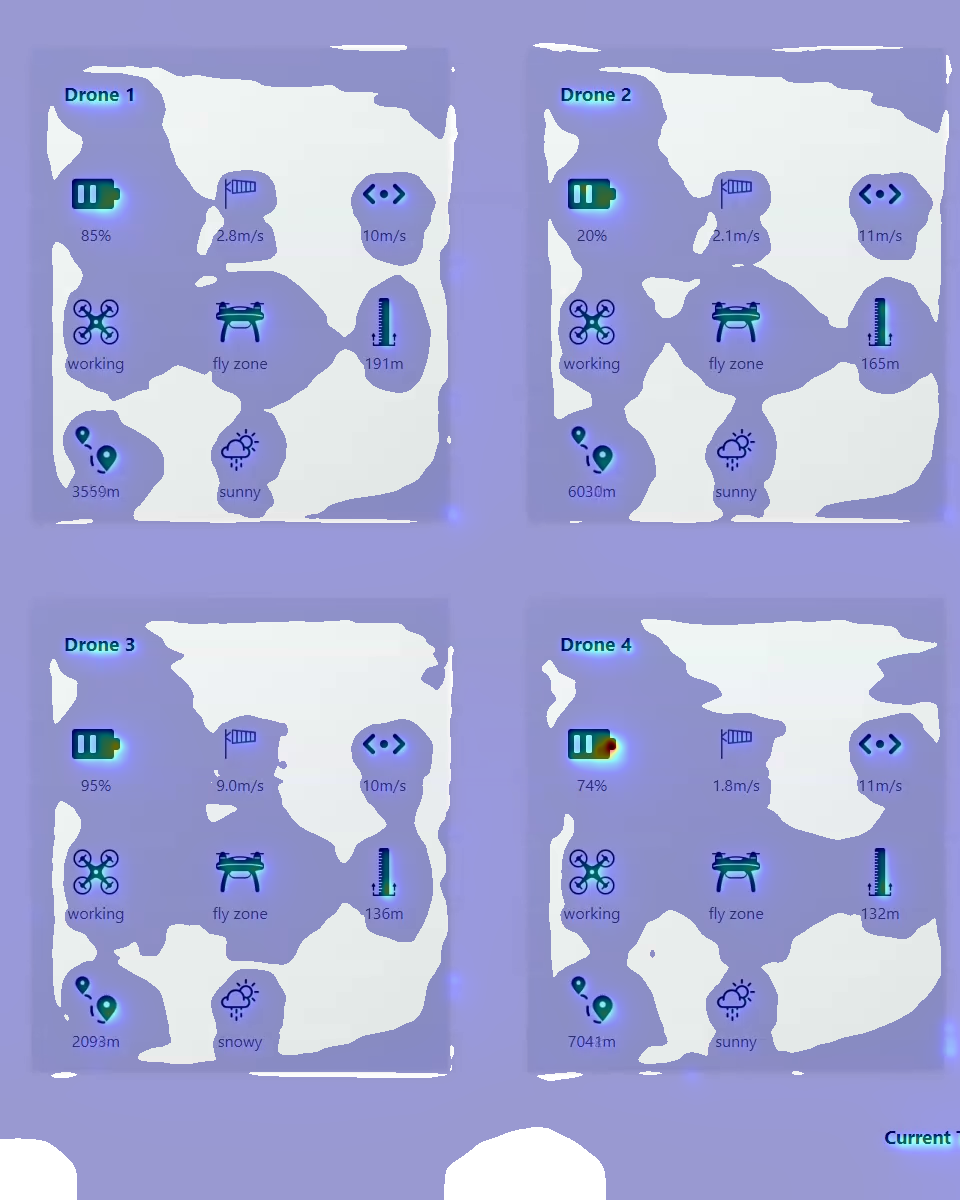}
  \end{subfigure}
  \hspace{0.5em}
  \begin{subfigure}{.2\textwidth}
    \includegraphics[width=\linewidth]{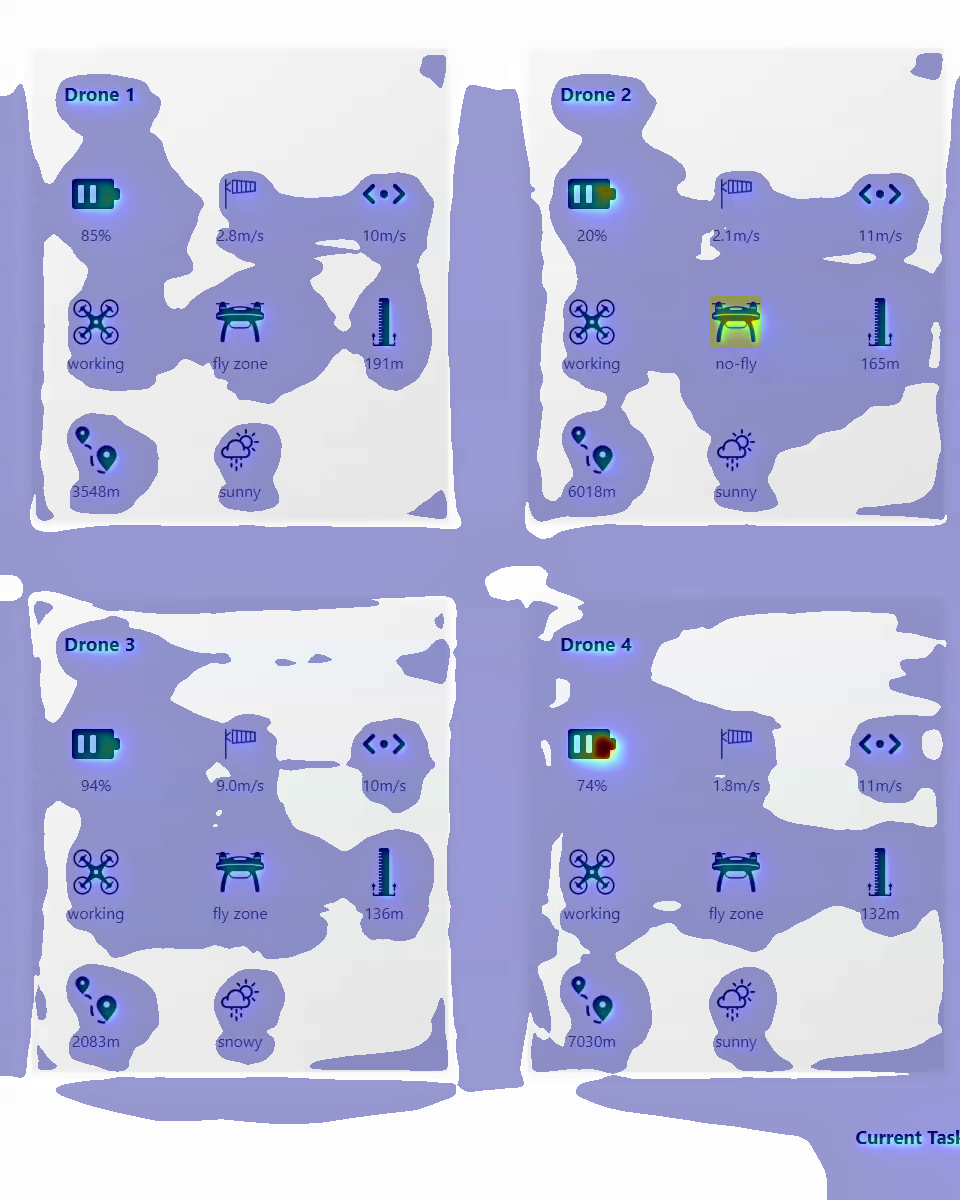}
  \end{subfigure}
  \hspace{0.5em}
  \begin{subfigure}{.2\textwidth}
    \includegraphics[width=\linewidth]{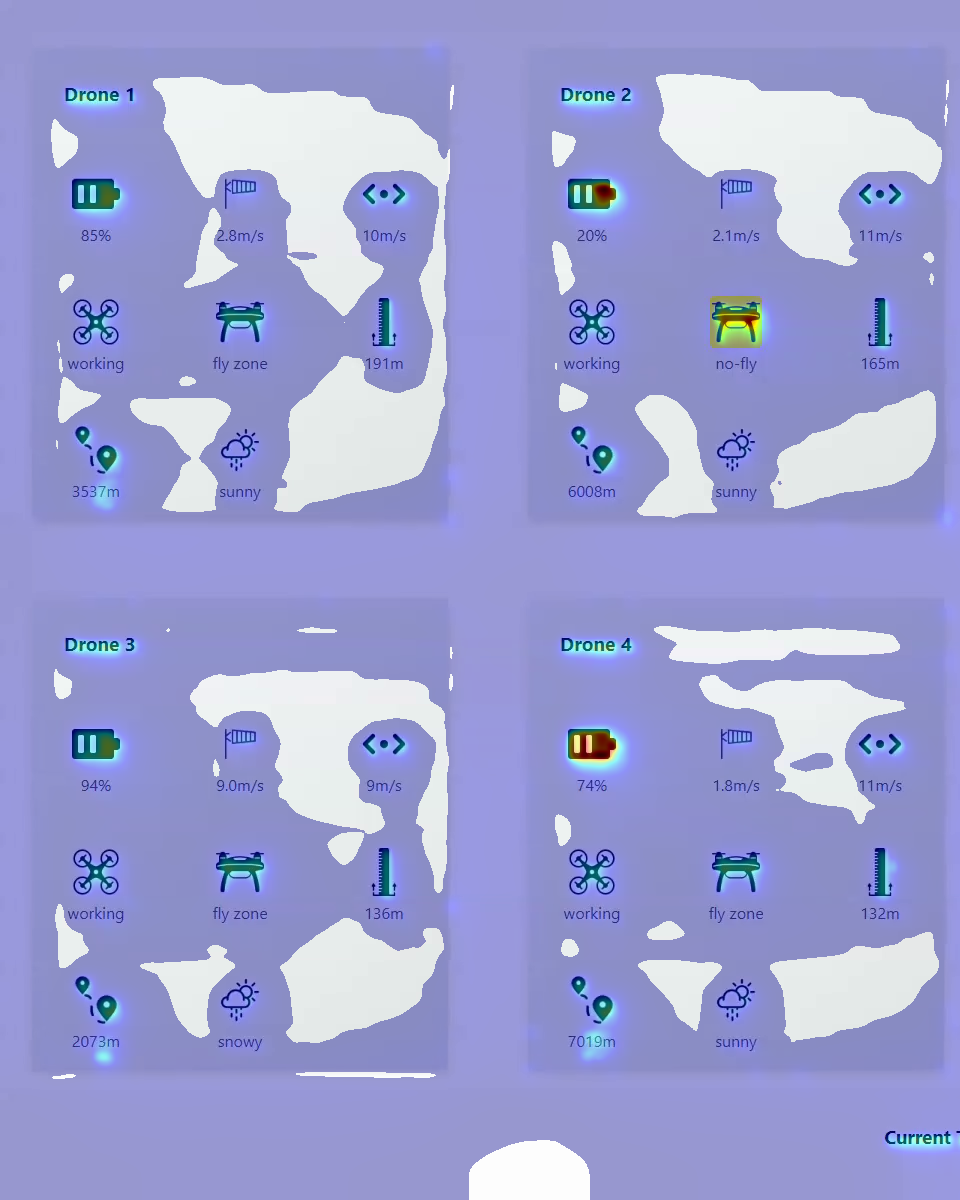}
  \end{subfigure}
  \hspace{0.5em}
  \begin{subfigure}{.2\textwidth}
    \includegraphics[width=\linewidth]{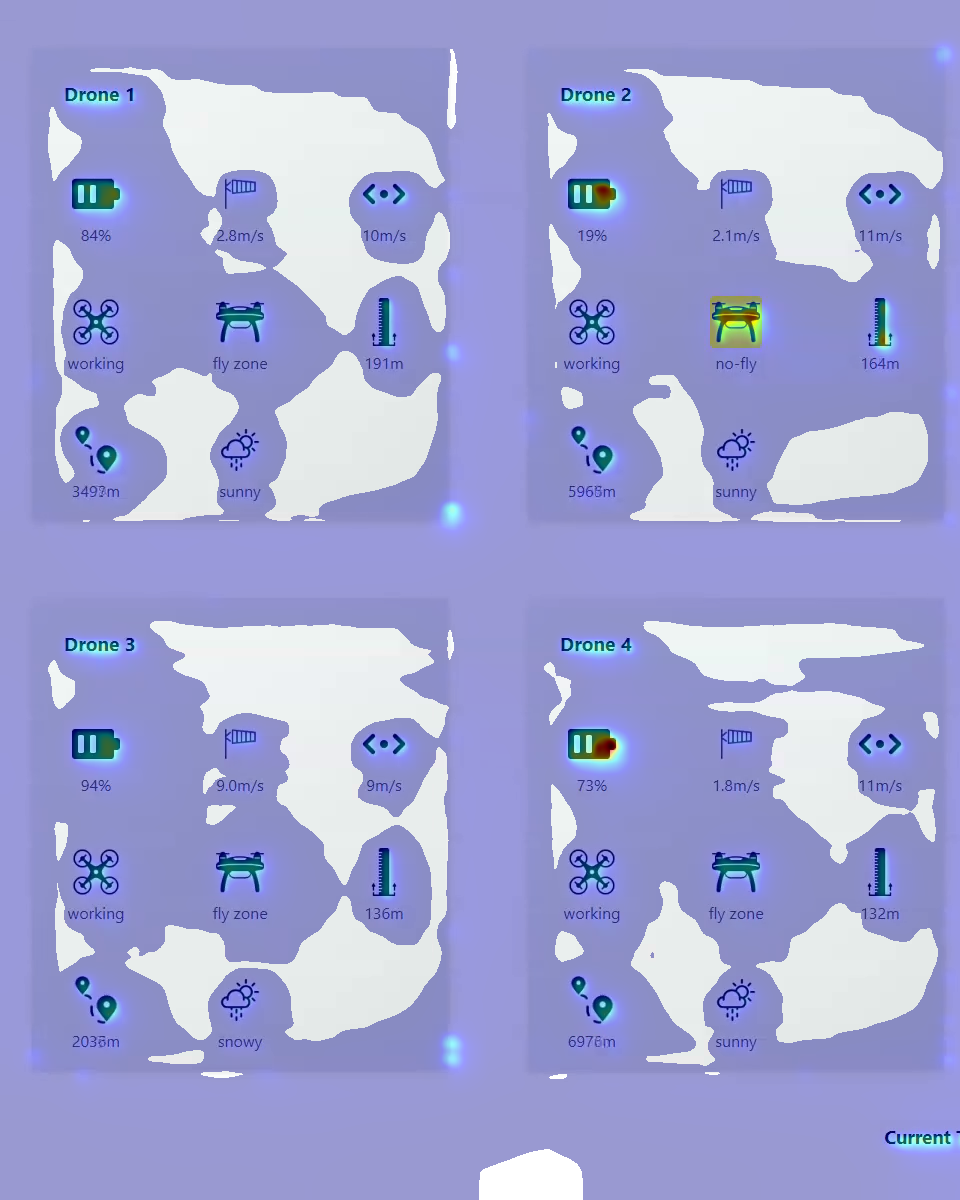}
  \end{subfigure}
\end{minipage}

\vspace{0.5em}

\begin{minipage}{.05\textwidth}
\raggedright
\textbf{SN}
\end{minipage}%
\begin{minipage}{.95\textwidth}
  \begin{subfigure}{.2\textwidth}
    \includegraphics[width=\linewidth]{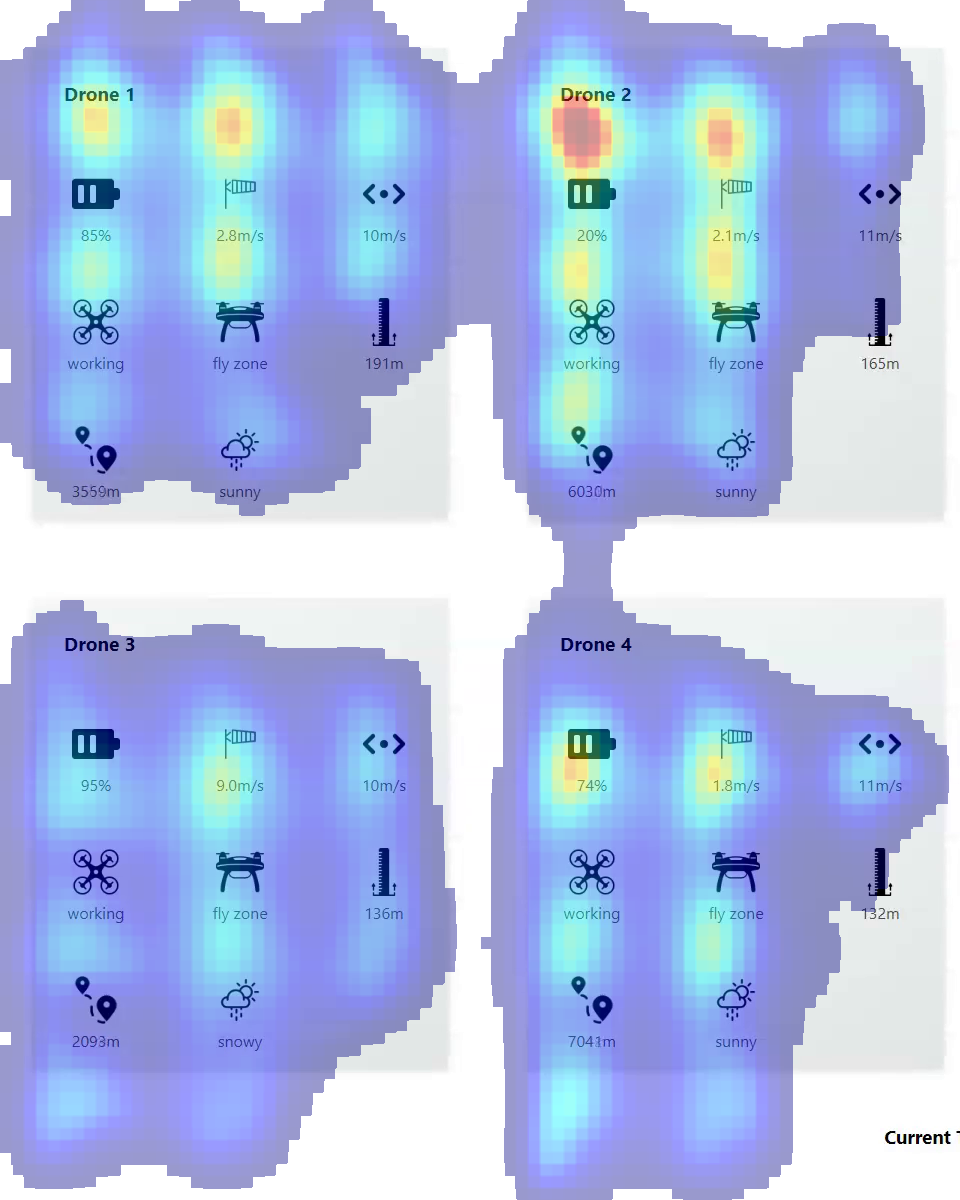}
  \end{subfigure}
  \hspace{0.5em}
  \begin{subfigure}{.2\textwidth}
    \includegraphics[width=\linewidth]{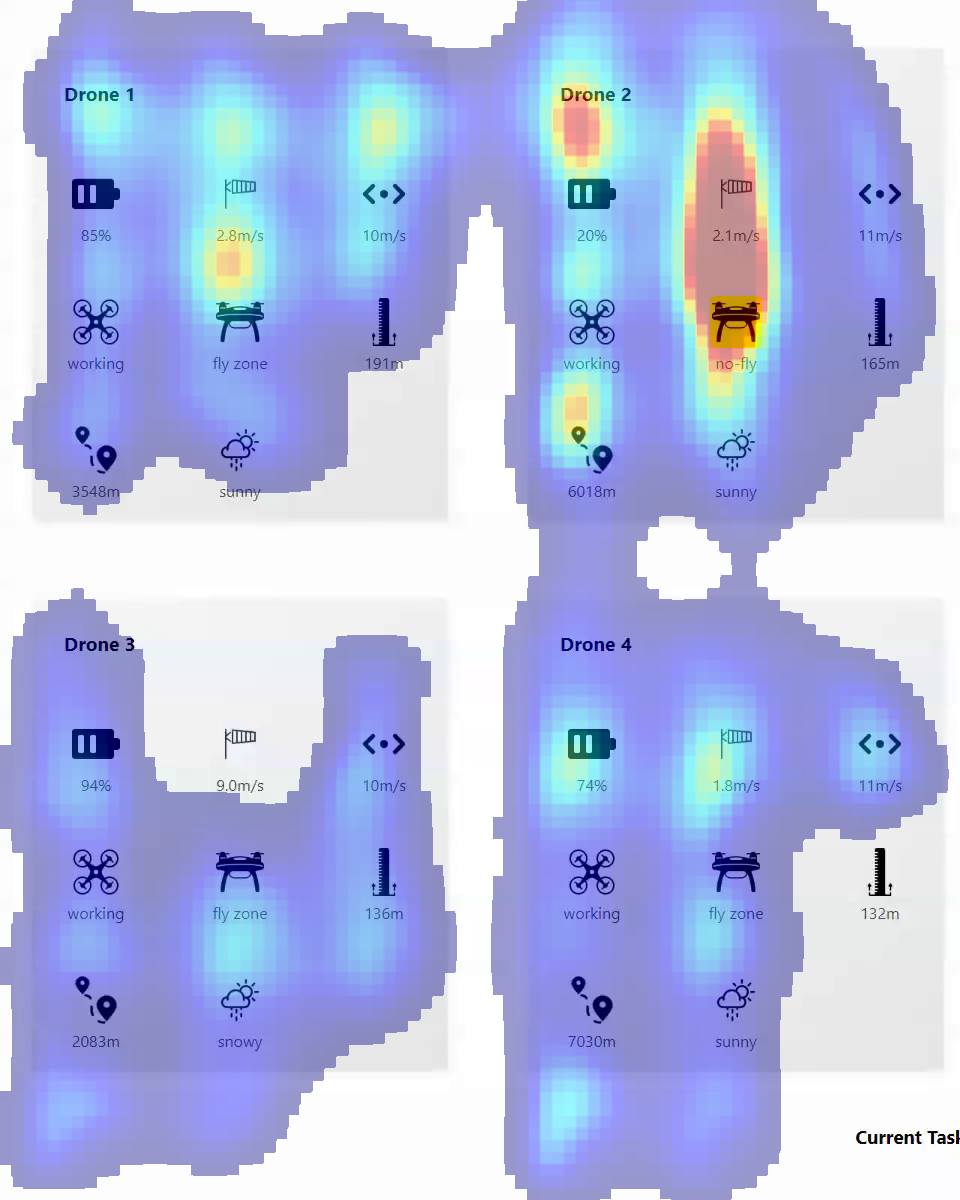}
  \end{subfigure}
  \hspace{0.5em}
  \begin{subfigure}{.2\textwidth}
    \includegraphics[width=\linewidth]{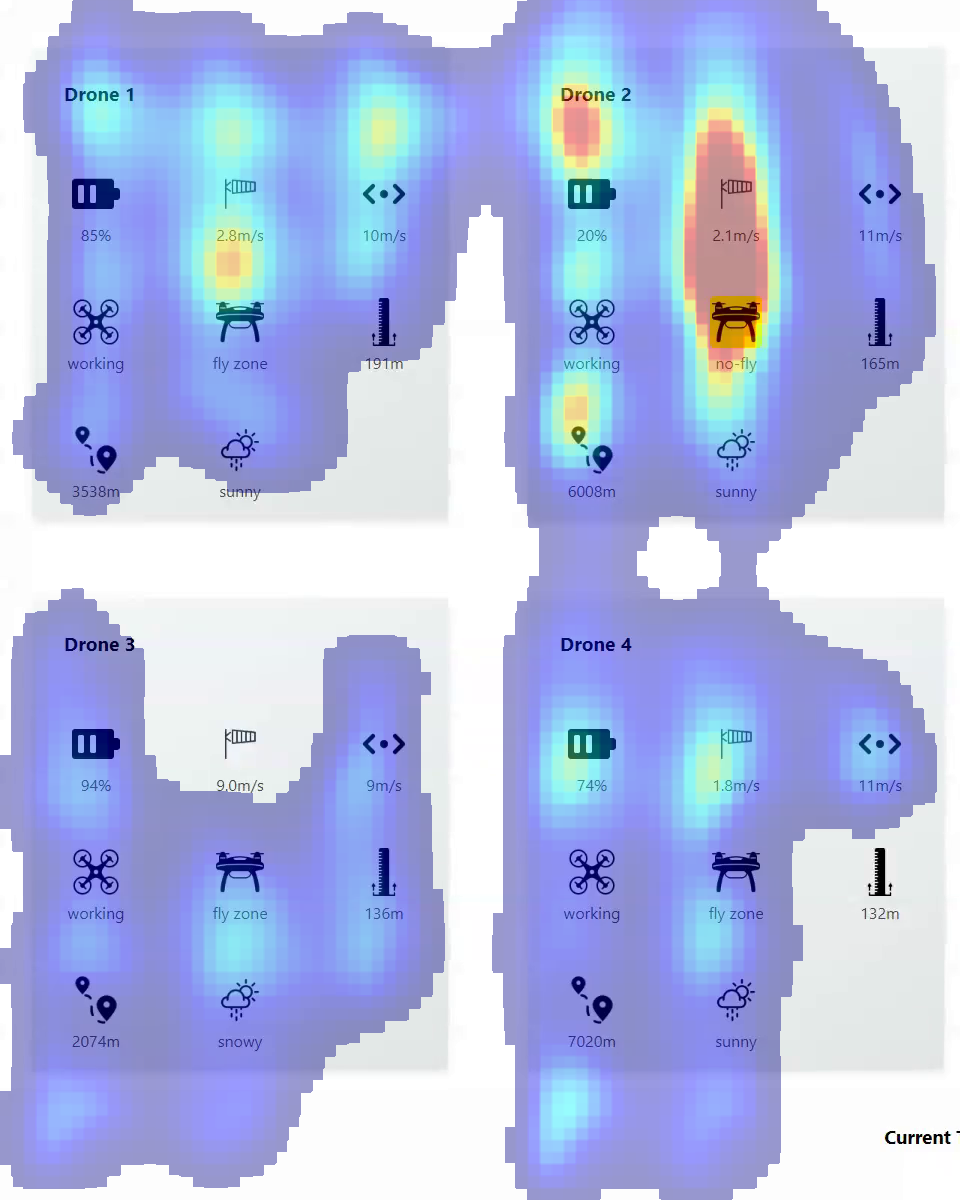}
  \end{subfigure}
  \hspace{0.5em}
  \begin{subfigure}{.2\textwidth}
    \includegraphics[width=\linewidth]{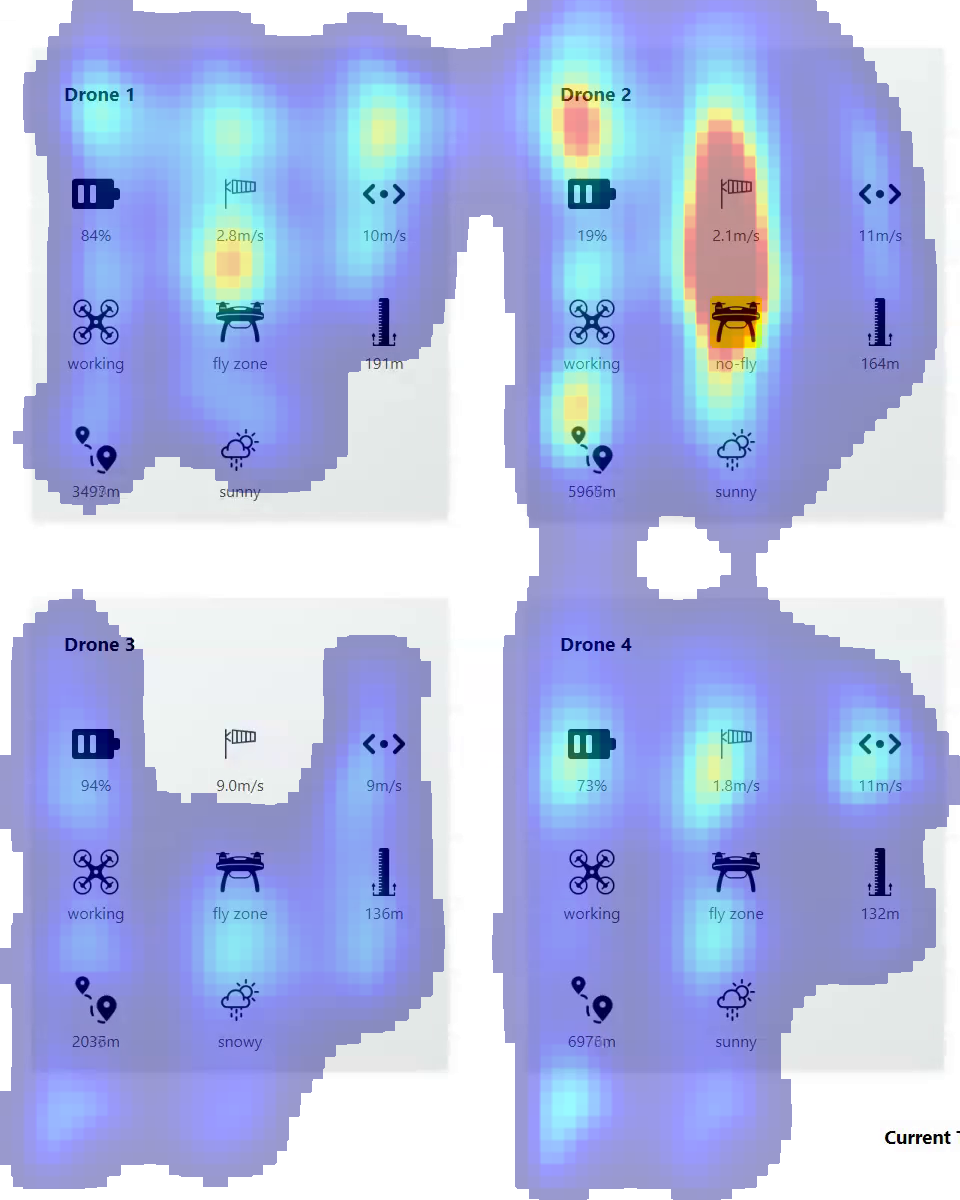}
  \end{subfigure}
\end{minipage}

\vspace{0.5em}

\begin{minipage}{.05\textwidth}
\raggedright
\textbf{TN}
\end{minipage}%
\begin{minipage}{.95\textwidth}
  \begin{subfigure}{.2\textwidth}
    \includegraphics[width=\linewidth]{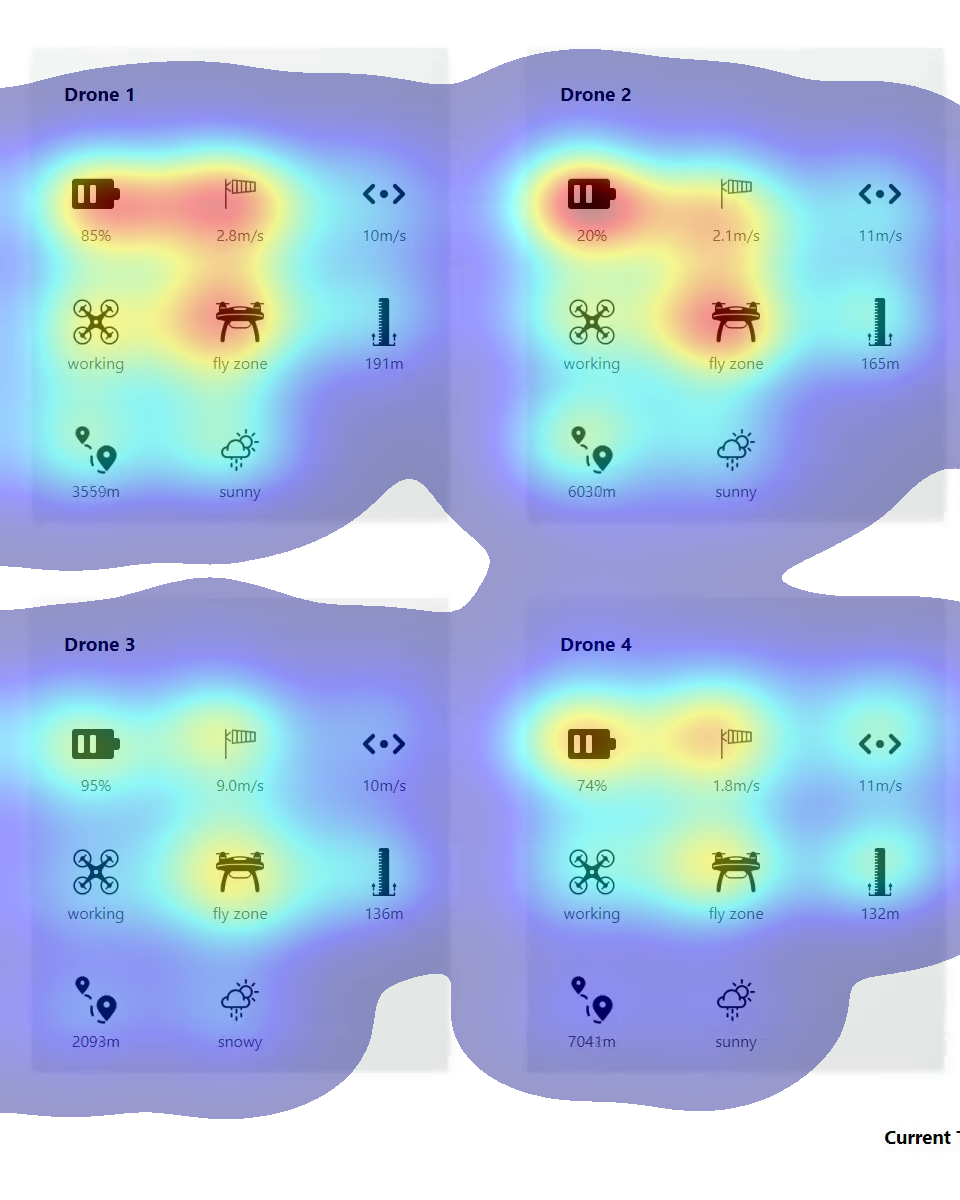}
    \caption*{T0 (1s Before CS)}
  \end{subfigure}
  \hspace{0.5em}
  \begin{subfigure}{.2\textwidth}
    \includegraphics[width=\linewidth]{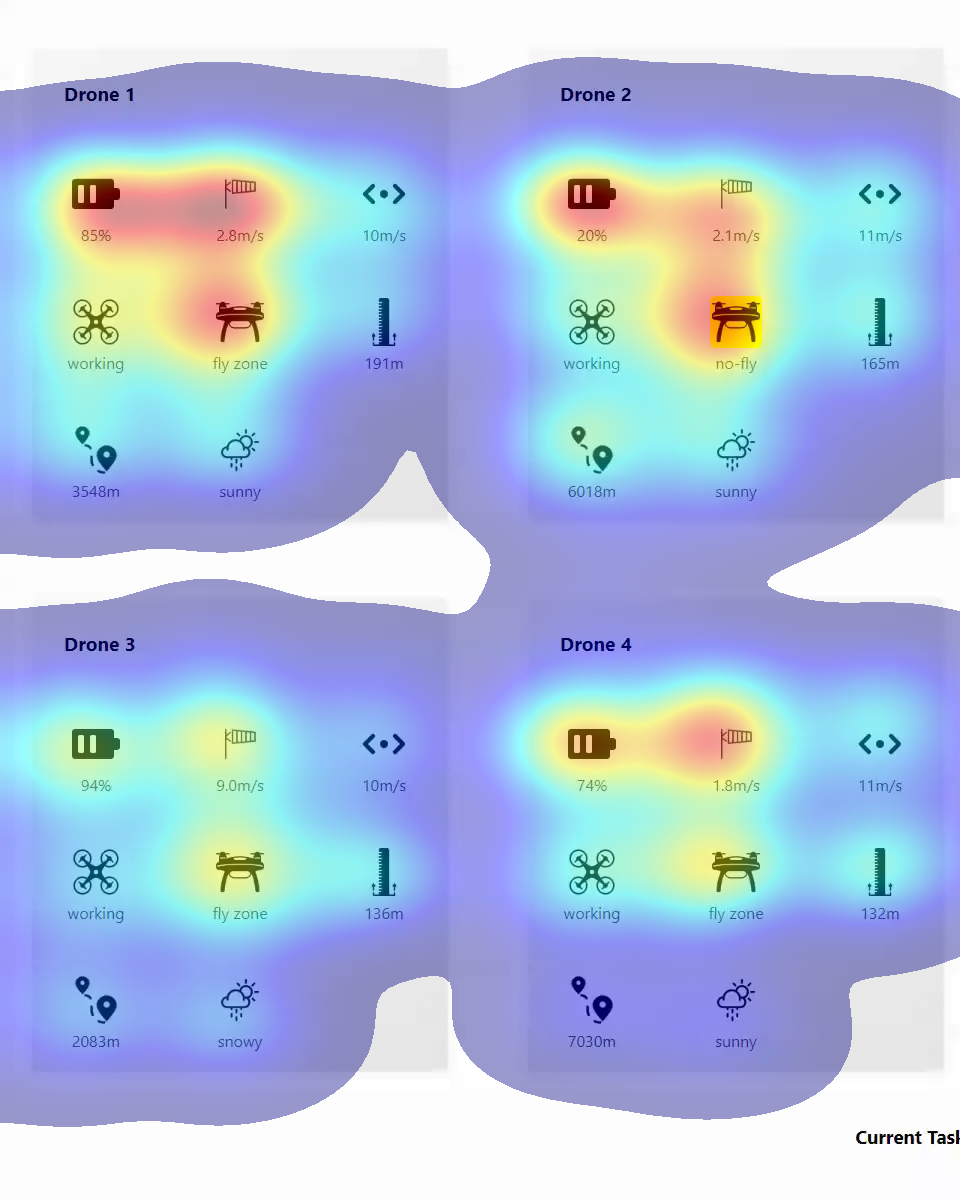}
    \caption*{T1 (CS Happened)}
  \end{subfigure}
  \hspace{0.5em}
  \begin{subfigure}{.2\textwidth}
    \includegraphics[width=\linewidth]{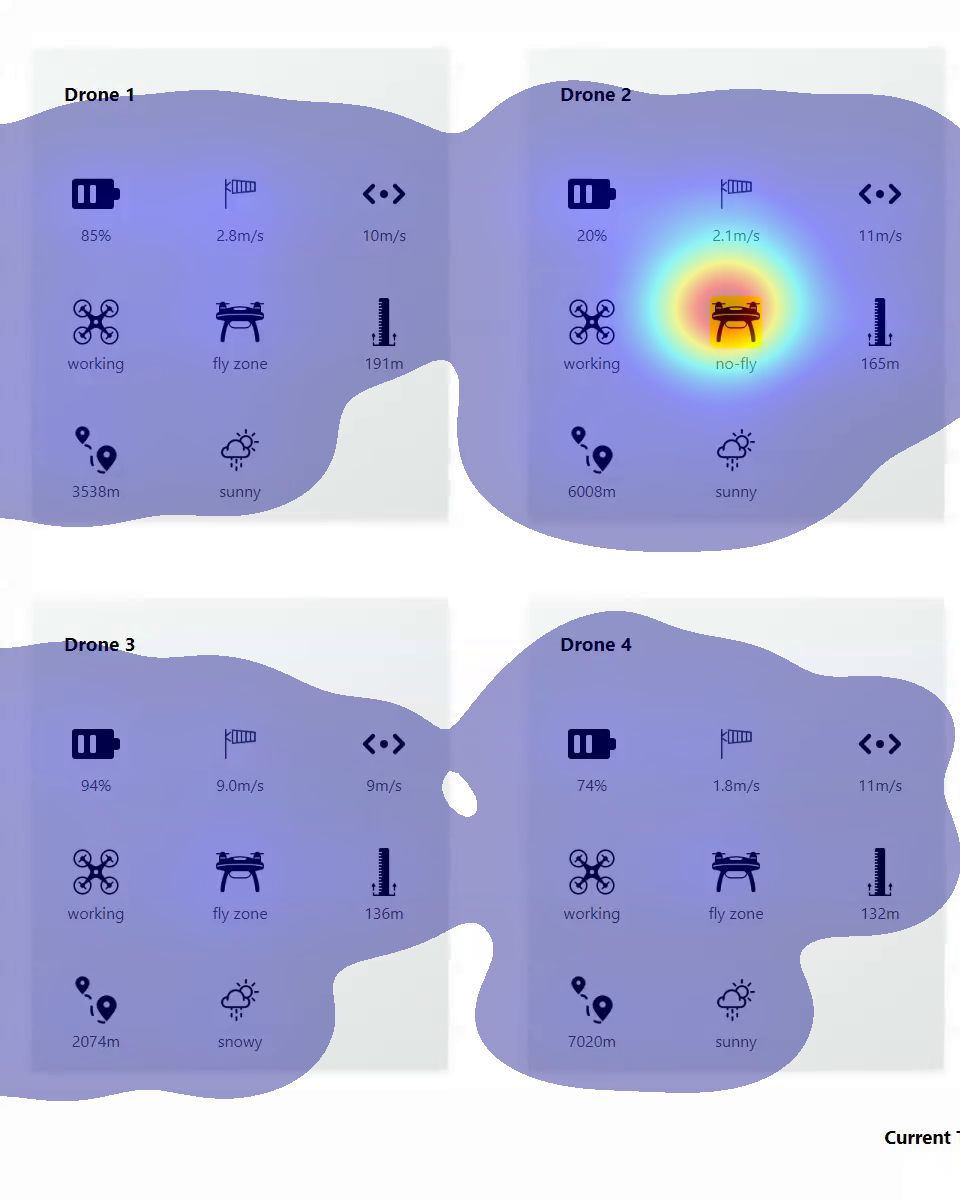}
    \caption*{T2 (1s After CS)}
  \end{subfigure}
  \hspace{0.5em}
  \begin{subfigure}{.2\textwidth}
    \includegraphics[width=\linewidth]{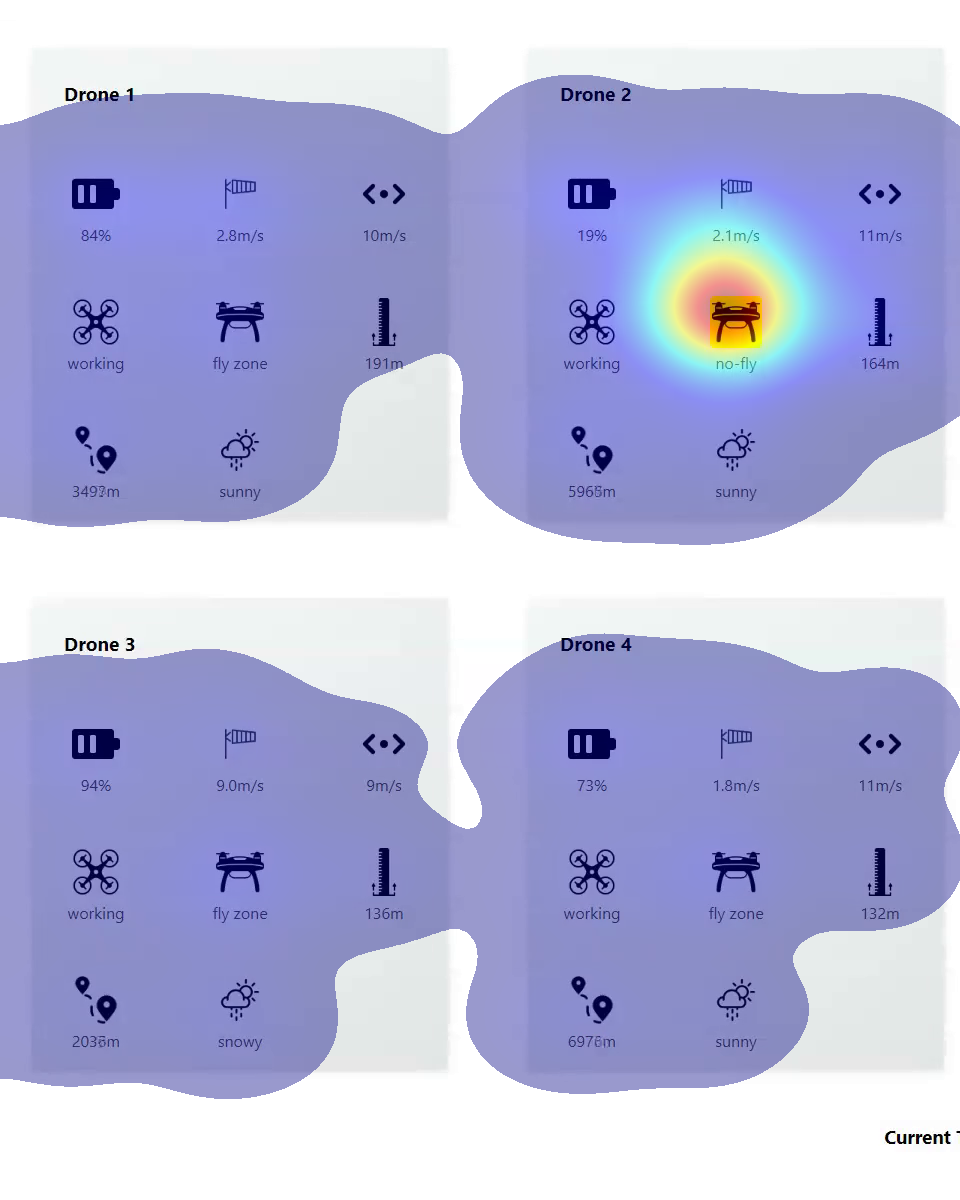}
    \caption*{T3 (5s After CS)}
  \end{subfigure}
\end{minipage}

\caption{Saliency model predictions for a highlighted critical situation (CS). 
The top row shows ITTI (IT), the middle row shows SimpleNet (SN), and the bottom row shows TASED-Net (TN).}
\label{fig:QA_high}
\end{figure*}

\subsubsection{Results}

\begin{table}
    \centering
    \renewcommand{\arraystretch}{0.9} 
    \setlength{\tabcolsep}{4pt} 
    \caption{Saliency Map Prediction Results under Different Highlight Conditions (Mean)}
    \label{tab:SP_HH}
    \begin{tabular}{l@{\hskip 6pt}c@{\hskip 6pt}c@{\hskip 6pt}c@{\hskip 6pt}c@{\hskip 6pt}c}
        \toprule
        Model & Highlight & AUC $\uparrow$ & SIM $\uparrow$ & NSS $\uparrow$ & CC $\uparrow$ \\
        \midrule
        \revision{ITTI} & H  & 0.484 & 0.140 & 0.740 & 0.139 \\
                        & NH & 0.461 & 0.142 & 0.467 & 0.107 \\
        \midrule
        SimpleNet & H  & \textbf{0.914} & \textbf{0.570} & \textbf{2.874} & \textbf{0.717} \\
                  & NH & \textbf{0.896} & \textbf{0.569} & \textbf{1.913} & \textbf{0.651} \\
        \midrule
        TASED-Net & H  & 0.876 & 0.451 & 2.166 & 0.535 \\
                  & NH & 0.860 & 0.475 & 1.645 & 0.511 \\
        \bottomrule
    \end{tabular}
\end{table}

\revision{\autoref{tab:SP_HH} presents the performance metrics of the three saliency models under two conditions: when a critical situation was highlighted and when it was not. Across all models, performance declined in the no-highlight condition, though the extent of the drop varied. Notably, there is a clear performance gap between the traditional ITTI model and the data-driven approaches—SimpleNet and TASED-Net—with both learning-based models achieving substantially better results. Among them, SimpleNet slightly outperformed TASED-Net across all quantitative metrics.}

\revision{However, a different picture emerges in the qualitative analysis of how well each model captures attention shifts triggered by highlights. As shown in \autoref{fig:QA_high}, ITTI produces dispersed saliency maps across all timestamps, failing to emphasize any specific interface elements. SimpleNet, while achieving higher metric scores, generates nearly identical predictions across highlighted frames (T1, T2, T3), as it processes each frame independently and lacks temporal awareness. In contrast, TASED-Net demonstrates a promising capacity to incorporate temporal context. Its predictions clearly shift from broadly distributed attention at the onset of the critical situation (T1) to a concentrated focus on the highlighted icon one second later (T2), closely mirroring the actual attention dynamics observed in \autoref{fig:visual_attention_temp}.}

Although TASED-Net is able to reflect the initial delay in detecting highlighted GUIs elements, there are several key limitations in using pixel-level saliency models to analyze how visual highlights modulate user attention over time. First, pixel-level saliency metrics provide a static, frame-by-frame evaluation of prediction accuracy but fail to capture how attention dynamically shifts in response to visual highlights. While SimpleNet achieves higher metric scores in isolated evaluations, its predictions remain largely invariant across timestamps, failing to reflect the gradual transition of visual focus. Second, these models do not allow for a direct comparison of the relative saliency of the highlight against other critical visual elements, such as drone icons, making it difficult to assess the user's SA within the broader interface context. Finally, crucial temporal aspects—such as how quickly a highlight captures attention and how long this attention is sustained—cannot be directly inferred from pixel-level saliency maps, further limiting their ability to characterize the dynamic effects of visual highlights. To address all these limitations, we transition to temporal normalized saliency prediction in the next section.

\subsection{Temporal Normalized Saliency Prediction}



\subsubsection{Task Description}


\revision{We formally define the \textit{temporal normalized saliency prediction task} as follows. Given a \textit{spatial input} $S' \in \mathbb{R}^{h \times w \times c}$, where $S'$ is a stacked image combining the global interface view $S^g$ and the local highlight region $S^e$, and \textit{temporal inputs} $v_t \in \{-1, 0, 1\}^T$ and $c_t \in \mathbb{R}^T$, where $v_t$ is a binary highlight vector indicating the presence (1), absence (–1), or padding (0) of highlights over the past $T$ time steps, and $c_t$ encodes the corresponding drone state values for the targeted icon—  
the task is to predict the \textit{output} $\hat{\text{NS}}(e, t) \in [0, 1]$, representing the normalized saliency of the highlighted element $e$ at timestamp $t$.}

Formally, the model learns the following mapping:
\[
f : (S', v_t, c_t) \mapsto \hat{\text{NS}}(e, t)
\]
The model is trained to minimize the discrepancy between the predicted value $\hat{\text{NS}}(e, t)$ and the ground truth $\text{NS}(e, t)$, enabling accurate temporal modeling of attention dynamics driven by visual highlights.



\subsubsection{HISM Model Structure}

  
 \revision{HISM consists of two main components: a \textit{spatial branch} and a \textit{temporal branch}, as shown in Figure~\ref{fig:model}. The spatial branch processes a stacked image that combines the global interface view and the local highlight region, allowing the model to capture spatial characteristics such as local contrast and layout context.}

 \revision{To capture temporal attention dynamics, we implemented \textbf{three variants} of the temporal branch:
\begin{itemize}
  \item \textbf{HISM (LSTM)}: uses an LSTM \cite{graves2012long} to process the highlight vector, which encodes the presence, absence, or padding of highlights over the past time steps.
  \item \textbf{HISM (TranEnc)}: replaces the LSTM with a Transformer Encoder \cite{vaswani2017attention}, using the same highlight vector as input.
  \item \textbf{HISM (TranEnc + TaskVec)}: extends the Transformer Encoder variant by concatenating the highlight vector with a task vector that encodes the drone state values of the targeted icon at each timestamp.
\end{itemize}}

 \revision{The task vector introduces top-down information that complements the highlight signal and helps the model make accurate predictions, especially in the absence of visual cues.}


\begin{figure}[t]
    \centering
    \includegraphics[width=\columnwidth]{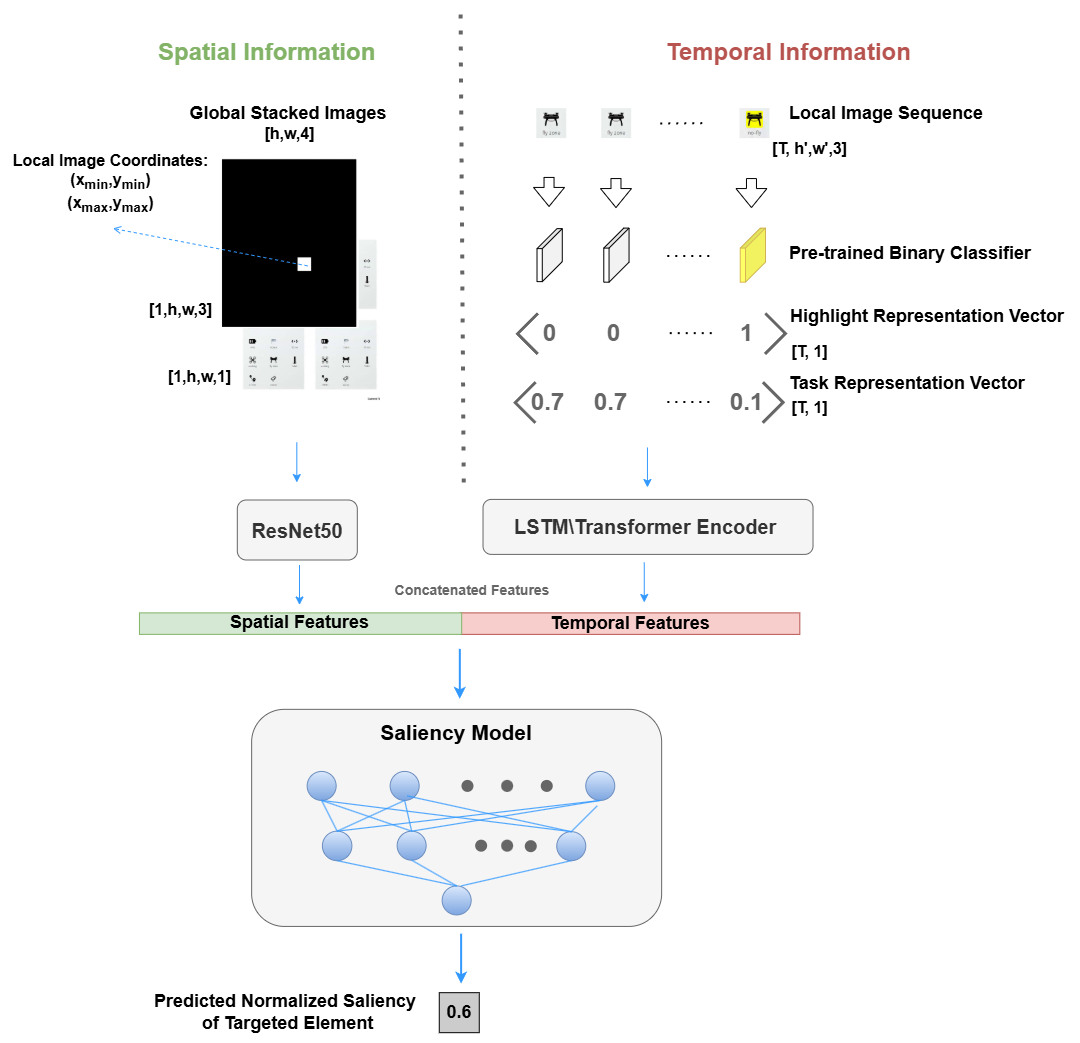}
    \label{fig:model}

    \caption{The overall architecture of the HISM.}
    \label{fig:model}
\end{figure}


The architecture of HISM (illustrated in \autoref{fig:model}) includes the following key steps:

\begin{enumerate}
    \item \textbf{Spatial Processing}: The stacked image $S'$, containing both the global GUI and the binary mask of the highlighted region, is processed through a ResNet50 backbone pre-trained on saliency prediction tasks \cite{huang2015salicon}. The spatial features are extracted via a global average pooling layer.

    \item \textbf{Temporal Processing}: \revision{We implemented three temporal variants: one using an LSTM and two using Transformer Encoders. While all variants take the highlight vector, $v^t$, as input, the final variant also incorporates a task vector, $c_t$, representing the drone state at each time step. These alternatives allow us to compare the impact of model architecture and input richness on attention prediction performance}.

    \item \textbf{Feature Fusion and Prediction}: The spatial and temporal features are concatenated and passed through a fully connected neural network consisting of three layers. These layers use ReLU activations and dropout regularization to predict the NS ($\hat{\text{NS}}(e, t)$) of the highlighted UI element for the given time slice. Since NS is a continuous scalar value, the model is optimized using the Mean Squared Error (MSE) loss function, which minimizes the difference between predicted and ground truth NS values. (Detailed training procedures are provided in the supplemental material.)

\end{enumerate}

This dual-branch architecture allows HISM to effectively leverage both the static spatial properties and the dynamic temporal characteristics of highlights. By integrating these components, HISM captures the evolving saliency of a UI element, enabling accurate prediction of how and when a highlight attracts attention. 




\subsubsection{Baselines \& Evaluation Metrics}
To assess the performance of the HISM model on the temporal NS prediction task, we compared its prediction results with the ground-truth data and the results generated from the \revision{three previous pixel-level saliency models: ITTI, SimpleNet, and TASED-Net}. As none of these models is designed to directly predict the NS of the highlighted element, we first predict the saliency map for the corresponding timestamps and then calculate the NS for the targeted element using \autoref{eq:normalized_saliency}. 

We evaluated the models using two metrics: \textit{MSE (Mean Squared Error)}, which measures the average squared difference between predicted and ground-truth values, with lower values indicating better predictions; and \textit{MAE (Mean Absolute Error)}, which calculates the average absolute difference between predicted and ground-truth values, providing a direct measure of prediction accuracy.

\subsubsection{Results}

\begin{table}
    \centering
    \renewcommand{\arraystretch}{0.9} 
    \setlength{\tabcolsep}{6pt} 
    \caption{Model Performance Evaluation under Different Highlight Conditions (Mean)}
    \label{tab:drone_results}
    \begin{tabular}{l@{\hskip 6pt}c@{\hskip 6pt}c@{\hskip 6pt}c}
        \toprule
        \textbf{Model} & \textbf{Highlight} & \textbf{MSE} $\downarrow$ & \textbf{MAE} $\downarrow$ \\
        \midrule
        \revision{ITTI} & H  & 0.0534 & 0.1502 \\
                        & NH & 0.0047 & 0.0515 \\
        \midrule
        SimpleNet       & H  & 0.0411 & 0.1255 \\
                        & NH & 0.0028 & 0.0407 \\
        \midrule
        TASED-Net       & H  & 0.0302 & 0.1145 \\
                        & NH & 0.0054 & 0.0589 \\
        \midrule
        HISM(LSTM)      & H  & 0.0062 & 0.0558 \\
                        & NH & 0.0030 & 0.0400 \\
        \midrule
        \revision{HISM(TranEnc)} & H  & 0.0052 & 0.0527 \\
                                 & NH & 0.0037 & 0.0521 \\
        \midrule
        \revision{HISM(TranEnc+TaskVec)} & H  & \textbf{0.0048} & \textbf{0.0472} \\
                                         & NH & \textbf{0.0022} & \textbf{0.0345} \\
        \bottomrule
    \end{tabular}
\end{table}

\revision{As indicated in \autoref{tab:drone_results}, we observe several notable patterns in model performance on the temporal normalized saliency (NS) prediction task. First, the highlight condition poses greater prediction challenges compared to the no-highlight condition. This is due to the dynamic nature of attention under highlighting—NS rises sharply to a peak and then quickly declines—introducing more temporal variance. In contrast, NS remains relatively low and stable in the absence of highlights, making it easier to predict. As a result, all models show higher prediction errors under the highlight condition, unlike the pixel-level saliency task, where highlighting improved performance.}

\revision{Second, under the highlight condition, all three HISM variants—\textbf{HISM (LSTM)}, \textbf{HISM (TranEnc)}, and \textbf{HISM (TranEnc + TaskVec)}—consistently outperform the baseline models (ITTI, SimpleNet, and TASED-Net), demonstrating the advantage of directly modeling temporal saliency with appropriate architectural support. However, in the no-highlight condition, this advantage diminishes: SimpleNet surpasses both HISM (LSTM) and HISM (TranEnc), likely due to the lower temporal complexity in these cases.}

\revision{Third, incorporating the task vector provides a substantial performance boost. \textbf{HISM (TranEnc + TaskVec)} achieves the lowest error in both conditions, highlighting the value of combining highlight signals with task-relevant information. This suggests that modeling both bottom-up (highlight-driven) and top-down (task-driven) cognitive processes leads to more accurate attention prediction. Paired t-tests confirm that HISM (TranEnc + TaskVec) performs significantly better than all other models in the highlight condition for both MSE and MAE ($p < .05$).}



\begin{figure*}[t]
\centering
\begin{subfigure}{0.48\textwidth}
  \includegraphics[width=\linewidth]{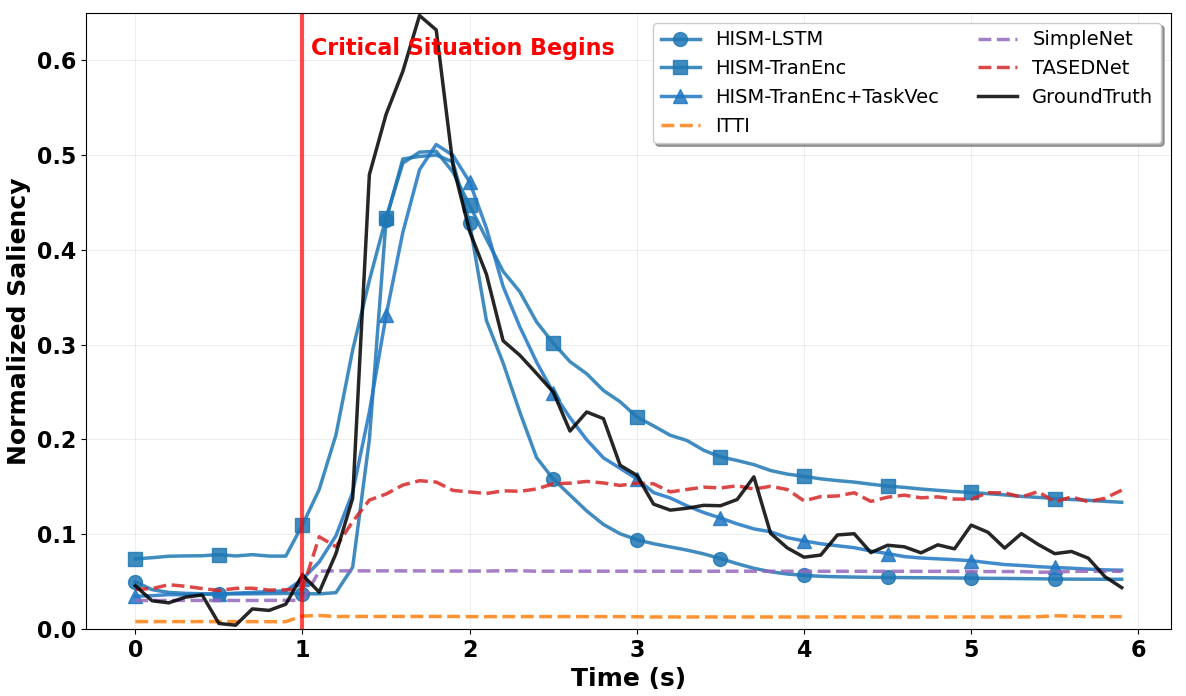}
  \caption{Highlight}
  \label{fig: baseline2}
\end{subfigure}
\hfill
\begin{subfigure}{0.48\textwidth}
  \includegraphics[width=\linewidth]{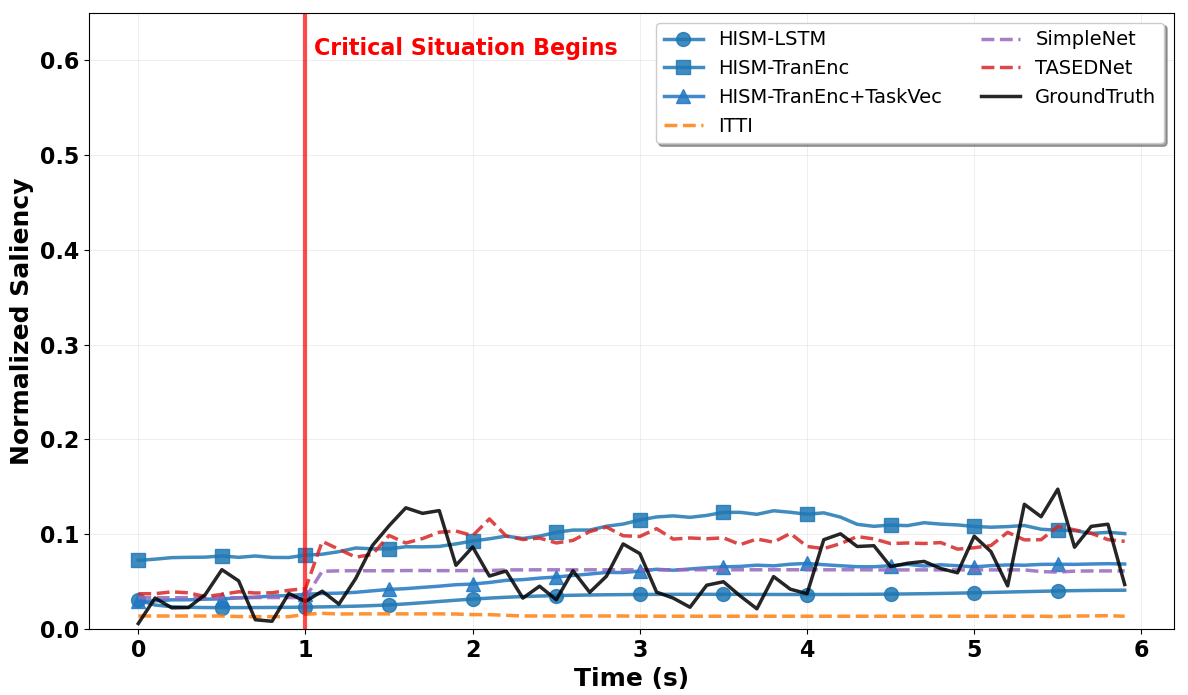}
  \caption{No Highlight}
  \label{fig: baseline3}
\end{subfigure}
\caption{The average prediction results of NS on the targeted element from our HISM-LSTM, HISM-TranEnc, HISM-TranEnc+TaskVec, the ITTI, the SimpleNet, and the TASED-Net on the test dataset (a) Highlight. (b) No Highlight.}
\label{fig:baseline}
\end{figure*}

\revision{Finally, the prediction curves (see Figure~\ref{fig:baseline}) further validate these findings. Under the highlight condition (Figure~\ref{fig: baseline2}), all HISM variants accurately capture the sharp rise in NS, peaking around 0.5 within the first second. In contrast, SimpleNet and TASED-Net show only minor changes and fail to reflect the peak or subsequent decline. Under the no-highlight condition (Figure~\ref{fig: baseline3}), NS changes are more random, yet all HISM variants still track a slight increase after the critical situation. Among them, HISM (TranEnc + TaskVec) yields the most accurate predictions.}


\section{Discussion \& Conclusion} 
\label{sec:conclusion}

In this paper, we investigated how visual highlights influence user attention in monitoring tasks, helping users focus on critical information more efficiently and accurately. To quantify the temporal effects of visual highlights, we introduced NS as a metric that captures when, how much, and for how long a highlighted element draws attention. Our findings revealed that NS not only reflects user engagement with the highlighted information at different timestamps but also provides insights into overall attention, which is closely related to SA. We proposed a new saliency model, HISM, to directly predict the NS of highlighted elements by integrating temporal and spatial information. HISM outperforms traditional pixel-level models by accurately inferring the NS of highlighted elements, enabling precise modeling of user attention dynamics.


\subsection{Supporting Human Oversight with Adaptive Interfaces}

Visual highlighting is widely used to direct attention efficiently, but it comes with potential trade-offs, especially in multitasking environments. While our findings reinforce that highlights can accelerate critical information detection, they also reveal unintended side effects, such as reduced engagement with other elements on the interface, which can impact SA. This aligns with concerns seen in alarm floods, where excessive cues lead to cognitive overload and reduced user oversight. Our study highlights the need to balance
attention guidance and interface adaptivity to avoid narrow visual tunneling while ensuring efficient task performance. Future adaptive interface designs should incorporate models like HISM to dynamically adjust visual cues based on user engagement patterns, ensuring that highlights support rather than hinder broader SA.

\subsection{Limitations and Future Work}

While our study focused on yellow highlights, the HISM framework is expected to generalize to a broader range of visual cues, including color changes, blinking effects, and shape modifications. \revision{However, we recognize that such abrupt onset cues may appear rigid or intrusive in certain safety-critical contexts. Future work should therefore not only investigate how various highlighting techniques influence attention distribution and task performance, but also explore more ecologically grounded alternatives. Inspired by Hansen’s work~\cite{hansen2018representation} on configural displays and optical invariants, promising directions include the use of \emph{figural goodness}, \emph{dynamic figural deformation}, and \emph{container-based representations}—approaches that subtly guide attention by leveraging perceptual principles rather than sudden visual changes. These alternatives could offer more cognitively compatible ways to direct user attention while maintaining broader situational awareness. Incorporating such mechanisms into predictive frameworks like HISM may enhance both usability and robustness across real-world monitoring interfaces.} \revisionsecond{At the same time, we acknowledge that HISM currently predicts the normalized saliency (NS) of a single element within a short detection window, while extending predictions to the full distribution of attention across the interface and to longer temporal horizons remains an open challenge. Moreover, our evaluation was conducted with 28 university participants; future work could examine expert populations and incorporate richer gaze signals, such as saccade sequences, to better capture domain-specific attentional strategies.}

\bibliographystyle{IEEEtran}
\bibliography{template}

\end{document}


\title{Supplementary Material: Detailed Research Components}
\author{}
\maketitle



\section{Calibration and Data Filtering}

The calibration involved a series of four calibration check pages, each devoted to one of the critical situation icons: battery, wind, rotor, and no-fly zone. On each page, a specific critical icon (for instance, the battery icon on the first page) would sequentially be highlighted in each of the four drone blocks, for a duration of 5 seconds each. Participants were asked to focus their attention on the highlighted icon in each drone block before moving on to the next. This sequential highlighting occurred first in the drone block one, followed by drone block two, then three, and finally, four, completing one calibration check page. This process was repeated across subsequent pages, with a different critical situation icon being highlighted in each page, thus ensuring that participants were familiarized with all critical icons across all drone blocks before commencing the main tasks of the study.

To sieve out sub-optimal data with poor tracking quality, we utilized the gaze data collected in the calibration part. Following previous research~\cite{feit2017toward}, we selected a 1-second window in which gaze points on average aligned closest with the target and computed the accuracy during that window for each of the four calibration pages (corresponding to the four tasks encountered by each participant).

We then excluded task data where the average offset in the x- or y-direction was larger than 70\,px from the highlighted icon center, corresponding to the mid-point between two icons in the same drone block. This resulted in the exclusion of 13\% of the initial gaze data.

\section{Participants Number}

\begin{figure}[h]
\centering
\includegraphics[width=0.95\columnwidth]{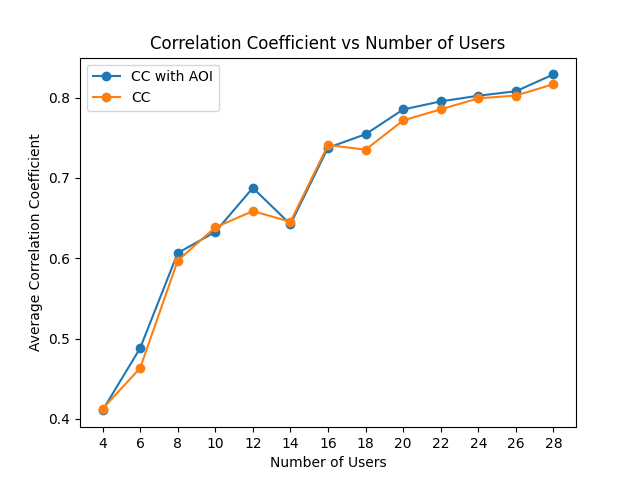}
\caption{The consistency of the recorded fixations for different numbers of participants. Our final dataset consisting of 28 participants shows a sufficient correlation in their gaze behavior.}
\label{fig:Fixation_CC}
\end{figure}

This method aimed to capture a diverse set of gaze behaviors while minimizing the impact of outliers. We randomly and evenly partitioned participants into two equal groups. This partitioning was replicated multiple times (e.g., five iterations). For each division, the linear CC between the fixation maps of the two groups was determined, averaging the CC values over the multiple iterations. The goal was to achieve convergence in the averaged CC value, indicating that participant recruitment could cease. As demonstrated in \autoref{fig:Fixation_CC}, given that the CC values for all fixations and AOIs areas stabilized above 0.8, our dataset comprising 28 participants is deemed sufficient. 

\section{Drone Simulation}

We developed a python script to simulate the drone parameters during the flight, which includes the drone flight route determination, the drone parameters updating, the critical situation handling, the data recording \& analysis and the question \& highlight flag.

\subsection{Drone Flight Route Determination}
The script initializes the drone's flight parameters including its starting and ending coordinates, which are chosen randomly within predefined latitude and longitude bounds. The trajectory of the drone and its distance to the destination are computed based on these coordinates, controlling the drone's flight route during the simulation.

\begin{itemize}
    \item Initial and destination coordinates are chosen randomly within set boundaries.
    \item The bearing and distance to the destination are calculated to determine the drone's trajectory.
    \item As the drone approaches its destination, its horizontal speed decreases linearly, and it starts descending when it's within 5\% of the initial distance.
\end{itemize}

\subsection{Updating Parameters Over Time}
As the simulation progresses, various drone parameters such as battery level, altitude, and speed are updated at each frame. The details are as follows:

\begin{itemize}
    \item Battery level decreases at specified intervals with a restriction to prevent dropping below a 10\% charge.
    \item Altitude and distance to destination are updated based on the drone's speed, which fluctuates with a certain noise.
    \item The distance to the destination diminishes steadily, considering the horizontal speed of the drone.
\end{itemize}

\subsection{Handling Critical Situations}
During its flight, the drone can encounter several critical situations that are handled by different functions modifying the corresponding parameters. The scenarios simulated are:

\begin{itemize}
    \item \textbf{Battery Drop}: Significantly decreases the battery value to simulate a sudden loss of power.
    \item \textbf{Extreme Wind}: Sets the wind speed to a high random value, affecting the drone's stability and speed.
    \item \textbf{Rotor Off}: Simulates a rotor malfunction, causing a continuous decline in altitude.
    \item \textbf{No-Fly Zone Warning}: Activates when the drone enters a predefined no-fly zone, altering the zone parameter accordingly.
\end{itemize}

\subsection{Data Recording and Analysis}
The script records the state of the drone at each frame, saving the data in a JSON file for subsequent analysis. Additional details include:

\begin{itemize}
    \item Detailed state information is saved at each time frame.
    \item A CSV file logs critical events and tasks, containing data on the occurrence of critical situations and other notable events.
\end{itemize}

\subsection{Question and Highlight Flags}
During the simulation, specific intervals are flagged for highlights or to trigger questions, promoting further analysis or user interaction. These flags serve as markers for critical or important events during the simulation. 

\begin{itemize}
    \item Highlight intervals signify periods where critical situations or special events occur.
    \item Question flags may prompt user interaction or further analysis at specific frames, though the exact nature is not detailed in the script.
\end{itemize}

\section{Iterative Design of the User Interface}

\subsection{First Iteration: Simulated Camera View}

\begin{figure}[h]
\centering
\includegraphics[width=0.95\columnwidth]{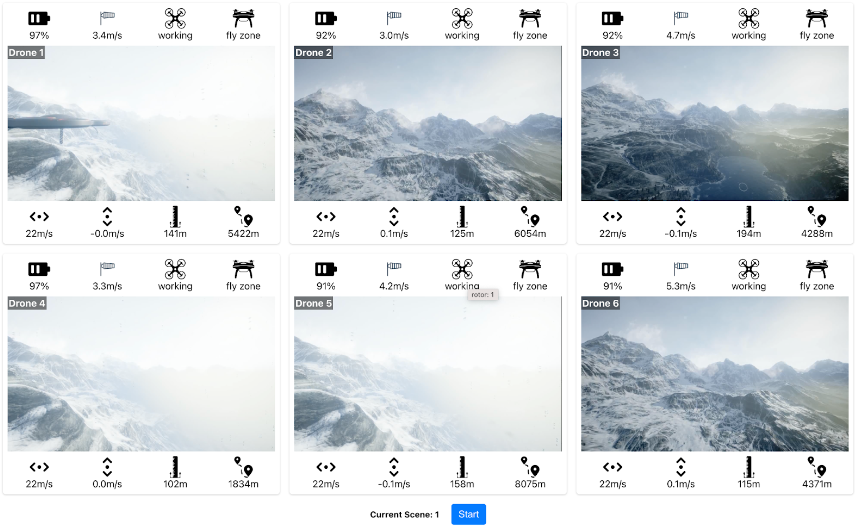}
\caption{Second Demo}
        \label{fig:demo1}
\end{figure}

As indicated in Fig. \ref{fig:demo1}, in the initial phase, a simulated camera view for each drone was integrated to craft a more immersive user experience. However, pilot tests highlighted that the inclusion of multiple video feeds escalated the cognitive load, adversely affecting task performance.

\subsection{Second Iteration: Icon-Based Interface}
Following this, an icon-centric approach was adopted to reduce visual complexity and cognitive demands on users. Despite simplifying the interface, this design rendered the display excessively static, potentially diminishing user engagement.

\begin{figure}[h]
\centering
\includegraphics[width=0.95\columnwidth]{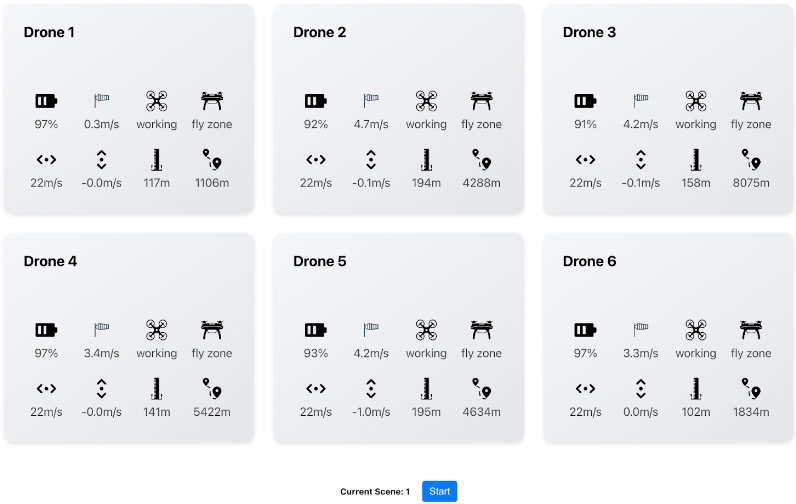}
\caption{Second Demo}
        \label{fig:demo2}
\end{figure}

\subsection{Final Iteration: Hybrid Approach}

\begin{figure}[h]
\centering
\includegraphics[width=0.95\columnwidth]{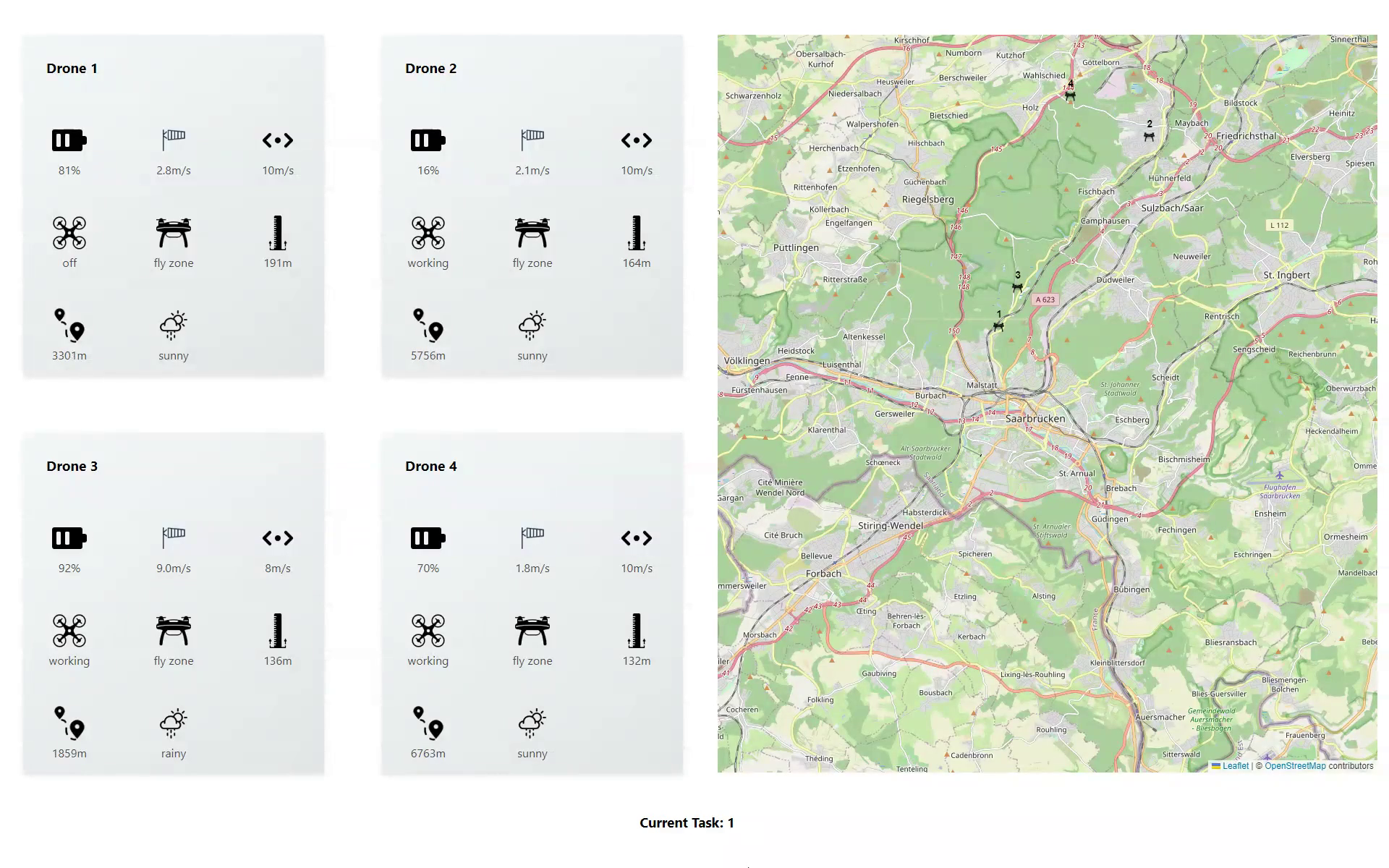}
\caption{Final Demo}
        \label{fig:GP}
\end{figure}

The final version combined positive aspects from the previous iterations (Fig. \ref{fig:GP}), incorporating iconography for rapid data assimilation and a map-based display to enhance spatial context. This hybrid approach improved user engagement and efficiency in drone monitoring tasks.

\section{Details in Saliency Model Training Process}
\subsection{Loss function choices for pixel-level saliency model}
The original SimpleNet was trained with KL and CC. We adapted the SimpleNet model by modifying its loss function to incorporate another SIM loss. This modification yielded an improved loss function formulation of 10KL-3CC-2SIM, reducing the occurrence of false predictions and better aligning with ground truth distributions.

The TASED-Net processes sequences of frames through a 3D convolutional encoder. Thus, this model requires a sequence of inputs (we chose a batch of 32 frames in this case), presenting a challenge given the limited size of our dataset. To navigate this, the encoder portion of the TASED-Net was frozen. We focused our efforts on fine-tuning the decoder section, optimizing the learning process within the constraints of our data set. Additionally, we adjusted the TASED-Net's loss function to a formulation of KL-0.5CC-0.1SIM, a modification supported by previous research findings to enhance the predictive accuracy and reliability of the model.

We used Python 3.9, with deep learning frameworks PyTorch v1.10.2 for both models and trained them on a machine equipped with an Intel i7-13700k CPU, 32GB RAM, and an NVIDIA RTX 4080 GPU.
The Adam optimizer, with a learning rate of $10^{-3}$ was used for training. 

\subsection{Training process for element-level saliencymodel}

To train the HISM with the global stacked image, $S^{'}$, we first resized both the global image and the mask into the size 300$\times$300 pixels. For the paired local image sequence, $v_t$, we included images showing the presence or absence of highlights based on their occurrence within the current time stamp $t$. These local images were then selected and padded into a sequence with a fixed length $T = 60$ covering the entire time duration. Before training, we defined the ground truth value as a single scalar representing the NS of the targeted icon at the time stamp $t$. In total, we had 1920 input pairs ($S^{'}$ and $Sq^{t}$) with ground truth values ((800 with highlight and 1120 without)). 

For training and testing the HISM, we took a similar approach as we did in the saliency map prediction to divide the dataset (60\% for training, 10\% for validation, and 30\% for testing). We use a batch size of 32 and an initial learning rate of 1e-4, reduced by a factor of 0.8 whenever the validation loss plateaued for five epochs. Early stopping was employed if the validation loss did not improve for ten epochs to avoid overfitting. The network was trained using Adam with MSE loss function. The project was implemented using Python 3.9. The computational resources included a machine equipped with an Intel i7-13700k CPU, 32GB RAM, and an NVIDIA RTX 4080 GPU.

\section{Post-Study Questionnaire}
Participants were asked to fill out the following questionnaire at the conclusion of the study to collect their feedback and perceptions:

\begin{enumerate}
  \item During the task, what factors did you consider when deciding which drone needed your attention?

  \textbf{Summarized Answers:} During the task, respondents employed various strategies to determine which drone necessitated their attention. A common approach was the circular or clockwise pattern of monitoring, aiming to allocate attention equitably amongst the drones. However, attention was predominantly guided by critical changes in certain parameters, including proximity to battery and wind speed thresholds, as well as other variables showing significant fluctuations. Several individuals emphasized the importance of closely monitoring drones that exhibited battery levels around 10\% or wind speeds nearing 9.9 m/s, as these figures signaled potential critical situations. Moreover, a switch in focus was noted if consecutive warnings were not perceived on the same drone, allowing for a broader oversight during the operation. This strategic redistribution of attention helped in maintaining a balanced view and in averting potential crises effectively.
  
  \item Did you notice any patterns or trends in the drone conditions that influenced your monitoring strategy?

 \textbf{Summarized Answers:} In response to the question about identifying patterns or trends in drone conditions that influenced their monitoring strategy, participants indicated a range of approaches. Many found altitude and velocity relatively stable and easier to monitor, thus requiring less attention after initial observations. Participants often focused on parameters that were approaching critical thresholds, shifting their focus to drones nearing 'border' conditions. Notably, after a critical event, there was usually a time delay before the next critical situation arose, allowing participants to monitor other non-critical data during this interval. Some respondents employed strategies such as following yellow markings and memorizing patterns to efficiently track drones. Additionally, they would sometimes disregard drones that had already experienced failures, focusing instead on monitoring battery levels, which tended to decrease with a degree of regularity. Changing weather conditions posed a challenge, making the monitoring task more difficult. This range of strategies reflects participants' efforts to adapt and respond dynamically to the evolving conditions during the drone monitoring task.
  
  \item How did you manage the simultaneous monitoring of multiple drones?

   \textbf{Summarized Answers:}Participants managed the simultaneous monitoring of multiple drones through various strategies, which primarily involved the systematic allocation of their attention across different drone parameters. A majority employed strategies such as round-robin, clockwise, or cross pattern scanning to maintain a balance in monitoring. Some focused on critical measurements such as battery levels and wind speeds, promptly addressing potential risks and anomalies. Others created mental abbreviations and categorizations to quickly assess and remember the conditions of each drone, aiding in their decision-making process during critical situations. Quick eye movements, frequent shifts in gaze, and memorization of certain values were also common techniques adopted to keep track of the rapidly changing data. Despite the initial challenges, participants noted that as the task progressed, they were able to adapt and pick up patterns that allowed for more efficient monitoring, even though some aspects remained tedious. This adaptive approach was instrumental in managing the complexity of monitoring multiple drones concurrently.
   
  \item From 1 (not at all) to 5 (very much), how much did you rely on the highlighting for monitoring the drones?
  
 \textbf{Visualized Results:} as indicated in Fig. \ref{fig:Q4}.
 
\begin{figure}[h]
\centering
\includegraphics[width=\columnwidth]{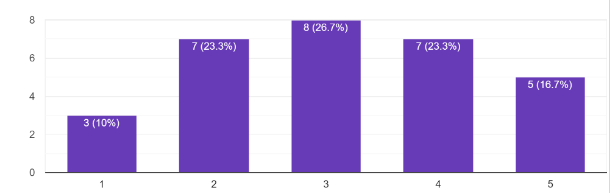}
\caption{}
\label{fig:Q4}
\end{figure}

  \item From 1 (not at all) to 5 (very much), how much did you trust the highlighting to correctly indicate a critical situation?
  
 \textbf{Visualized Results:} as indicated in Fig. \ref{fig:Q5}.
 
\begin{figure}[h]
\centering
\includegraphics[width=\columnwidth]{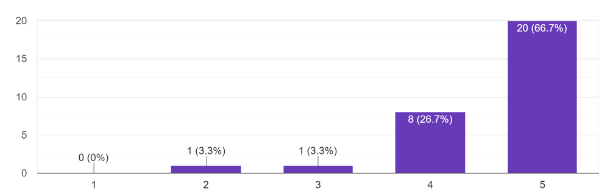}
\caption{}
\label{fig:Q5}
\end{figure}

  \item From 1 (totally disagree) to 5 (totally agree), how much do you agree with the following statement: "The highlighting helped me to detect critical situations"?
  
   \textbf{Visualized Results:} as indicated in Fig. \ref{fig:Q6}.
   
\begin{figure}[h]
\centering
\includegraphics[width=\columnwidth]{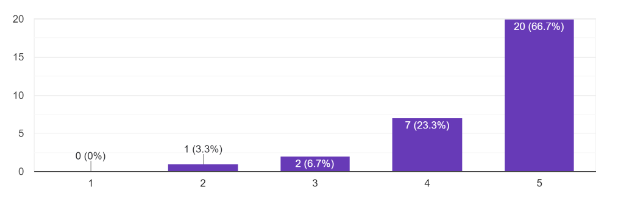}
\caption{}
\label{fig:Q6}
\end{figure}

  \item From 1 (very poor) to 5 (very good), how do you feel about your performance in identifying the critical situation?

   \textbf{Visualized Results:} as indicated in Fig. \ref{fig:Q7}.

\begin{figure}[h]
\centering
\includegraphics[width=\columnwidth]{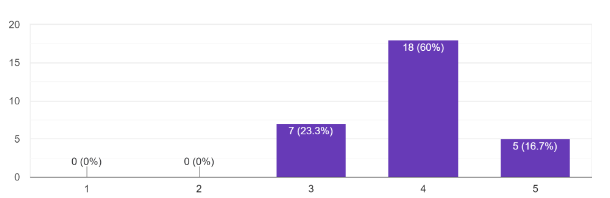}
\caption{}
\label{fig:Q7}
\end{figure}
  
  \item From 1 (very poor) to 5 (very good), how do you feel about your performance in completing the situation awareness questionnaire?

   \textbf{Visualized Results:} as indicated in Fig. \ref{fig:Q8}.

\begin{figure}[h]
\centering
\includegraphics[width=\columnwidth]{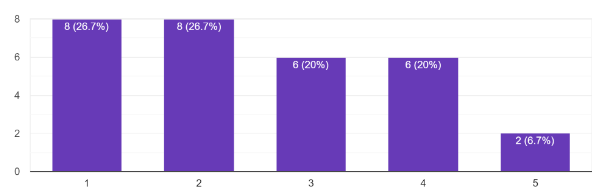}
\caption{}
\label{fig:Q8}
\end{figure}
  
  \item From 1 (not demanding) to 5 (very demanding), how mentally demanding was the task?

    \textbf{Visualized Results:} as indicated in Fig. \ref{fig:Q9}.

\begin{figure}[h]
\centering
\includegraphics[width=\columnwidth]{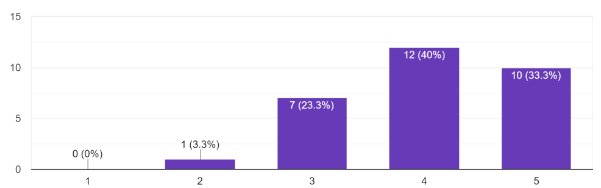}
\caption{}
\label{fig:Q9}
\end{figure}

  \item From 1 (not demanding) to 5 (very demanding), how physically demanding was the task?

    \textbf{Visualized Results:} as indicated in Fig. \ref{fig:Q10}.
  
\begin{figure}[h]
\centering
\includegraphics[width=\columnwidth]{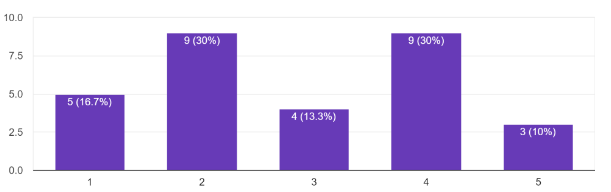}
\caption{}
\label{fig:Q10}
\end{figure} 
\end{enumerate}